\documentclass[a4paper,fleqn]{cas-dc}

\usepackage[utf8]{inputenc}
\usepackage{graphicx}
\usepackage{amsmath,amsfonts,mathtools,relsize}
\usepackage{vcell}
\usepackage{tabularray}
\usepackage{multirow}
\usepackage[numbers]{natbib}
\usepackage{pdflscape}
\usepackage{ragged2e}

\begin{document}

\let\WriteBookmarks\relax
\def\floatpagepagefraction{1}
\def\textpagefraction{.001}

\shorttitle{LATUP-Net: A Lightweight 3D Attention U-Net with Parallel Convolutions for Brain Tumor Segmentation}
\shortauthors{E.~Alwadee~\emph{et~al.}}

\title[mode=title]{LATUP-Net: A Lightweight 3D Attention U-Net with Parallel Convolutions for Brain Tumor Segmentation}

\author[1,2]{Ebtihal J. Alwadee}[orcid=0009-0004-3398-4023]
\ead{AlwadeeEJ@cardiff.ac.uk}
\credit{Conceptualization, Methodology, Software, Investigation, Writing - Original Draft}
\cormark[1]

\author[1]{Xianfang Sun}[orcid=0000-0002-6114-0766]
\ead{SunX2@cardiff.ac.uk}
\credit{Supervision, Writing - Review \& Editing}

\author[1]{Yipeng Qin}[orcid=0000-0002-1551-9126]
\ead{QinY16@cardiff.ac.uk}
\credit{Supervision, Writing - Review \& Editing}

\author[1]{Frank C. Langbein}[orcid=0000-0002-3379-0323]
\ead{frank@langbein.org}
\credit{Software, Data Curation, Supervision, Writing - Review \& Editing}

\affiliation[1]{organization={School of Computer Science and Informatics, Cardiff University},
    city={Cardiff},
    postcode={CF24 4AG},
    country={UK}}

\affiliation[2]{organization={Department of Computer Science, College of Engineering and Computer Science, Jazan University},
    city={Jazan},
    country={KSA}}

\cortext[cor1]{Corresponding author}

\begin{abstract}
Early-stage 3D brain tumor segmentation from magnetic resonance imaging (MRI) scans is crucial for prompt and effective treatment. However, this process faces the challenge of precise delineation due to the tumors' complex heterogeneity. Moreover, energy sustainability targets and resource limitations, especially in developing countries, require efficient and accessible medical imaging solutions. The proposed architecture, a Lightweight 3D ATtention U-Net with Parallel convolutions, LATUP-Net, addresses these issues. It is specifically designed to reduce computational requirements significantly while maintaining high segmentation performance. By incorporating parallel convolutions, it enhances feature representation by capturing multi-scale information. It further integrates an attention mechanism to refine segmentation through selective feature recalibration. LATUP-Net achieves promising segmentation performance: the average Dice scores for the whole tumor, tumor core, and enhancing tumor on the BraTS~2020 dataset are $88.41\%$, $83.82\%$, and $73.67\%$, and on the BraTS~2021 dataset, they are $90.29\%$, $89.54\%$, and $83.92\%$, respectively. Hausdorff distance metrics further indicate its improved ability to delineate tumor boundaries. With its significantly reduced computational demand using only $3.07$~M parameters, about $59$ times fewer than other state-of-the-art models, and running on a single NVIDIA GeForce RTX3060 12~GB GPU, LATUP-Net requires just $15.79$~GFLOPs. This makes it a promising solution for real-world clinical applications, particularly in settings with limited resources. Investigations into the model's interpretability, utilizing gradient-weighted class activation mapping and confusion matrices, reveal that while attention mechanisms enhance the segmentation of small regions, their impact is nuanced. Achieving the most accurate tumor delineation requires carefully balancing local and global features. The code is available at~\url{https://qyber.black/ca/code-bca}.
\end{abstract}

\begin{keywords}
  Brain Tumor Segmentation
  \sep Lightweight
  \sep Parallel Convolutions
  \sep Attention
  \sep U-Net
  \sep Deep Learning
  \sep Grad-CAM
\end{keywords}

\maketitle

\section{Introduction}

Brain tumors, particularly gliomas, are among the most lethal forms of cancer due to their inherent complexity and high variability among patients. Gliomas, classified into high-grade and low-grade, consist of different tumor regions, including the enhancing tumor, necrotic core, and surrounding edema~\cite{menze2014multimodal}. Magnetic resonance imaging (MRI) is the standard method for diagnosing gliomas, and precise segmentation of these tumors is crucial for effective treatment planning. Accurate segmentation helps clinicians differentiate tumor tissue from healthy tissue, which directly influences diagnosis, treatment strategies, and prognosis. However, manual delineation of tumor regions across MRI slices is both time-consuming and labor-intensive, with the accuracy highly dependent on the clinician’s expertise and subjective thresholding. This subjectivity, combined with the effort involved, underscores the need for efficient and accurate automatic segmentation techniques for brain tumors~\cite{icsin2016review}.

In this work we introduce a novel lightweight deep learning architecture, LATUP-Net (Lightweight 3D ATtention U-Net with Parallel Convolutions). It significantly reduces the computational resources needed while maintaining state-of-the-art brain tumor segmentation performance, as demonstrated on BraTS~2020~\cite{menze2014multimodal,bakas2017advancing,bakas2018identifying} and BraTS~2021~\cite{menze2014multimodal}.
We incorporate parallel convolutions in the first encoder block of a U-Net architecture, inspired by the inception block~\cite{li2019novel}. This reduces feature redundancy and parameter count by sharing an initial feature extraction stage across multiple convolutional paths, followed by pooling operations to capture multi-scale spatial features. The design harnesses diverse features efficiently while maintaining a lower parameter count than traditional inception blocks. We further add an extension mechanism, which generally yields smaller performance improvements, but also does not add substantial computational costs.

Deep learning, particularly convolutional neural networks (CNNs), has revolutionized medical image analysis by providing powerful tools for segmentation, classification, and object detection. CNNs are widely used in medical imaging due to their ability to automatically extract hierarchical features, making them well-suited for complex tasks like brain tumor segmentation~\cite{wieczorek2021lightweight, wozniak2021deep}. Among these architectures, U-Net has become a leading model for medical image segmentation and has demonstrated remarkable performance across various tasks. The U-Net architecture has been widely adopted, receiving over $20,000$ citations as of 2023, reflecting its profound impact on medical image segmentation research~\cite{ronneberger2015u,isensee2024nnu}. 
To meet the increasing demands of clinical applications, U-Net-based models have evolved by enhancing their network structures, incorporating new modules, and expanding to 3D networks~\cite{zhu2024sparse,xu2024brain,wu2022mr}. However, improving segmentation accuracy often comes at the cost of increased model complexity and parameters~\cite{sarvamangala2022convolutional}. Existing networks require large numbers of parameters and computational resources, which poses challenges for real-world use in resource-limited environments~\cite{beutel2000handbook}. To address this challenge, lightweight networks have been developed to reduce model parameters and computational resource requirements without compromising performance~\cite{wieczorek2021lightweight,ma2023lmu}. These models employ strategies such as full convolution and deep separable convolution to achieve high segmentation accuracy while minimizing resource usage, making them more suitable for deployment in resource-constrained environments. In our specific approach we exploit parallel convolutions to recuce computational resources requierd for training and inference.

In particular, in MRI scans with small tumor lesions, attention mechanisms have been employed to improve results to focus on a specific region and its features. Such integrating can improve the accuracy of small tumor segmentation~\cite{vaswani2017attention}. Common lightweight attention methods for medical image semantic segmentation include BAM~\cite{park2018bam}, CBAM~\cite{woo2018cbam}, and Squeeze-and-Excitation (SE)~\cite{hu2018squeeze}. BAM uses atrous convolution to achieve a larger receptive field, CBAM combines spatial and channel attention separately, and SE focuses on channel attention to address feature loss during convolutional pooling. However, using both spatial and channel attention increases computational complexity. The Squeeze-and-Excitation mechanism is chosen for its efficient feature extraction through channel attention while maintaining computational efficiency, making it suitable for small-scale segmentation tasks~\cite{hu2018squeeze}.

We explore the integration of attention mechanisms within LATUP-Net to improve segmentation of tumor regions. While attention mechanisms effectively emphasize tumor-specific features, our investigation highlights a trade-off: attention may focus too narrowly on these features and overlook broader contextual information, which is also essential for accurate segmentation. By balancing local detail and global context, LATUP-Net achieves performance comparable to state-of-the-art models while drastically reducing computational demands.

In summary, the contributions of our work are: 
\begin{itemize}
\item We introduce LATUP-Net, an efficient 3D U-Net variant that combines parallel convolutions and attention mechanisms to achieve high segmentation accuracy at a fraction of the computational cost of existing models.
\item We demonstrate the effectiveness of parallel convolutions in capturing multi-scale features, resulting in a richer, more efficient representation. They optimize feature diversity and computational efficiency by leveraging a shared initial convolution followed by distinct paths and pooling operations.
\item We critically examine the role of attention mechanisms in segmentation, using Grad-CAM~\cite{selvaraju2017grad} and confusion matrix analysis to assess their impact. Our findings show that while attention enhances focus on tumor features, a balanced approach considering both local detail and global context improves overall segmentation performance.
\end{itemize}

The subsequent sections of this paper are structured as follows: Section~\ref{sec:related} presents an overview of previous research conducted on the segmentation of brain tumors. Section~\ref{sec:architecture} explains the LATUP-Net architecture, encompassing its essential elements and their impact on performance. Section~\ref{sec:details} provides a comprehensive analysis of the experimental configuration, encompassing the datasets used and the evaluation metrics employed. Section~\ref{section:result} presents and analyses the obtained results. Finally, Section~\ref{sec:conclusion} concludes the paper and provides suggestions for further research.

\section{Related Work}\label{sec:related}

To analyze existing methods for brain tumor segmentation, we consider three perspectives: convolutional neural networks (CNN) and their variants, lightweight models, and attention mechanisms 

\subsection{CNN and Related Models}

Brain tumor segmentation using CNNs has been extensively studied in the literature, with most methods employing either 2D or 3D convolutions. Initially, 2D models dominated the field, where CNNs processed individual 2D slices derived from 3D MRI scans. However, these 2D slices inherently lack the volumetric context present in full 3D MRIs, leading to the potential loss of important semantic features. This issue is compounded by the fact that the resolution within the plane of 2D slices is often higher than that across slices, and the presence of small gaps between slices further exacerbates the loss of spatial continuity. To capture 3D feature information, 3D CNNs emerged as the preferred approach for analyzing MRI images of brain tumors, addressing the limitations inherent in 2D slice-based analysis~\cite{cciccek20163d}.

While 3D convolutions make better use of spatial information, they also require more computational power and memory. To address this issue, Chen~\emph{et al.}~\cite{chen2019s3d} developed a memory-efficient solution, while preserving most of the volumetric information, by introducing a decoupled 3D U-Net model. It relies on separating a 3D convolution into sequential 2D and 1D convolutions and creating three parallel branches of these separated convolutions, one for each orthogonal view (axial, sagittal, and coronal) for the 1D convolution direction. The suggested model achieved competitive results while demonstrating high efficiency when tested on the BraTS~2018 dataset. While local and global features are necessary for making decisions during segmentation, low-level feature gradients (such as those containing information about boundaries, edges, lines, or dots) converge to zero as one proceeds deeper into the model. To address this, Wang~\emph{et al.}~\cite{wang2021transbts} proposed a TransBTS architecture that effectively embeds a transformer into a 3D U-Net model. To begin with, local feature maps are extracted using a 3D CNN encoder. The extracted features are transmitted through a transformer to capture global features. Afterward, the decoder incorporates the local and global features during the upsampling process to produce the segmentation result. Zhu~\emph{et al.}~\cite{zhu2023brain} propose a BTS method that combines deep semantic features and edge features for semantic feature extraction and fusion from multimodal MRI. The method uses the Swin Transformer for semantic feature extraction, shifted patch tokenization for training efficiency, and an Edge Spatial Attention Block (ESAB) for feature enhancement. Though both models require more computational resources, it has shown promising results on BraTS~2018 to BraTS~2020, achieving competitive or higher Dice score performance compared with state-of-the-art 3D models.

Traditional U-Net architectures perform exceptionally well on semantic segmentation tasks~\cite{ronneberger2015u}. Nevertheless, such structures lack strategies to extract global feature information~\cite{liu2019automatic,oh2019automated}. To address this, the inception module~\cite{li2019novel} and a densely connected module~\cite{liu2019liver} were added to the U-Net architecture by Zhang~\emph{et al.}~\cite{zhang2020dense}. Each inception module in the network uses $1\times1$, $3\times3$, and $5\times5$ convolution kernels to acquire multi-scale information. Their method performs admirably in segmenting images of lung tissue, blood arteries, and brain tumors. Meanwhile, the transformers' self-attention mechanism automatically brings in global information but lacks the inductive bias, so it does not obtain sufficient fine-grained features. Thereby, combining transformers and CNNs may leverage the strengths of both. However, this combination often neglects lesion boundaries, which are essential for accurate segmentation. To address this issue, Xu~\emph{et al.}~\cite{xu2024brain} integrated the Swin-T network with a dual-path feature inference module to enhance the original Swin-T network, resulting in improved edge segmentation performance for cranial tumors. Zhu~\emph{et al.}~\cite{zhu2024sparse} used skip connections to integrate multi-level edge fusion features, derived from the sparse dynamic encoder, into the decoder, therefore improving the transmission of spatial edge information and further refining the network's segmentation performance.

While these approaches improve the representational power of the models, they also increase the number of parameters, heightening the risk of overfitting and limiting effectiveness in scenarios with limited training data. Therefore, given the constraints of low resource funding in many hospitals, it is crucial to strike a balance between processing efficiency and network size through the development of lightweight networks.

\subsection{Lightweight Models}

U-Net variants have demonstrated satisfactory segmentation results for medical images. However, 3D networks require significantly more GPU memory than 2D networks with the same convolutional network structure and depth. Consequently, hardware requirements limit improvements in segmentation. Researchers have proposed a series of lightweight models to reduce network complexity and overcome hardware limitations. By reducing the number of network parameters and achieving highly accurate segmentation, Chen~\emph{et al.}~\cite{chen20193d} created a dilated multi-fiber (DMF) network that replaces convolutions with dilated convolutions of varying sizes as the fundamental unit. Although dilated convolutions, which modify the convolution's field of view by introducing gaps in the convolutional kernel, can capture features at various scales, they do not necessarily increase the diversity of features captured. Luo~\emph{et al.}~\cite{luo2020hdc} proposed a lightweight hierarchical decoupled convolution (HDC) unit by replacing 3D convolutions with pseudo-3D convolutions. However, the model's final segmentation precision is not very good, despite its ability to explore multi-scale, multi-view spatial contexts rapidly with a large reduction in computing complexity. In addition, Magadza~\emph{et al.}~\cite{magadza2022brain} utilized depth-wise separable convolutions to reduce computational complexity without sacrificing performance. However, this method cannot handle diverse and fundamental features in multiple, independent directions and orientations.

In a recent study, Zhu~\emph{et al.}~\cite{zhu2024brain} proposed a CNN-based model for brain tumor segmentation. Their approach combines three modules that use multimodal, spatial, and boundary information. This method examined the overall spatial aspects of the image allowing it to accurately acquire the tumor's location and how it relates to other tissues in MRI scans. The proposed model proved to be more effective and efficient than other existing state of the arts methods.

In summary, while traditional 3D U-Net models and their variants offer high segmentation accuracy, they are often resource-intensive and prone to overfitting in limited data scenarios. In contrast, lightweight models like the DMF network~\cite{chen20193d}, HDC unit~\cite{luo2020hdc}, and depth-wise separable convolutions~\cite{magadza2022brain}, although less precise, provide a viable solution for environments with computational constraints. Our proposed lightweight 3D network addresses these concerns by employing a specific version of parallel convolutions which enhances feature extraction and segmentation performance with significantly fewer parameters, offering a balanced solution between computational efficiency and segmentation precision. We further incorporate an attention mechanism into our model, which addresses the overfitting issue typically associated with complex models.

\subsection{Attention Mechanism}\label{sec:ATT}

Traditional U-Nets give equal importance to all features within the feature maps. Given the notable class imbalance in brain tumor segmentation, some features are more crucial than others for accurate results. Attention mechanisms have emerged as effective tools to emphasize these crucial features and downplay less significant ones. Generally, attention mechanisms are bifurcated into two main types: channel attention and spatial attention. Channel attention enables the network to adaptively weigh the importance of different channels based on specific features in the image. This can potentially prioritize channels that are crucial for tumor detection~\cite{hu2018squeeze}. Spatial attention, instead, fine-tunes the spatial feature maps adaptively, allowing the network to concentrate on specific regions with significant features~\cite{roy2018concurrent}. In this study, we explore various lightweight attention mechanisms, all recognized for their capacity to enhance model expressiveness and boost overall performance.

\section{The LATUP-Net Architecture}\label{sec:architecture}

\begin{figure*}[t]
  \centering
  \includegraphics[width=.9\textwidth]{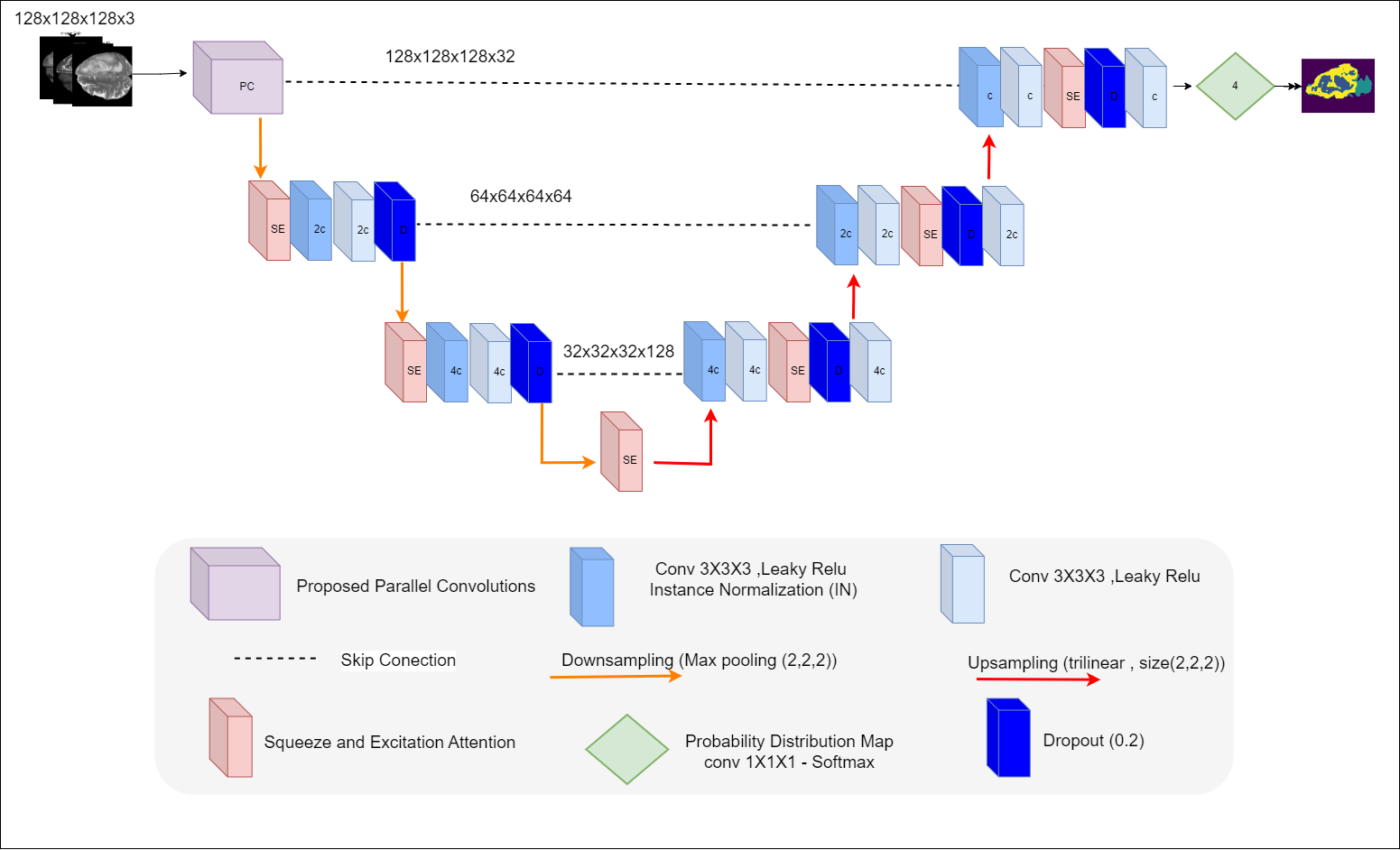}
  \caption{The LATUP-Net architecture.}\label{Figure:model}
\end{figure*}

Here, we explain the components of our LATUP-Net architecture, illustrated in Figure~\ref{Figure:model} and Table~\ref{tab:model_architecture}, a lightweight variant of the original U-Net~\cite{cciccek20163d} with fewer parameters intended for the semantic segmentation of 3D brain tumors. Moreover, LATUP-Net utilizes multi-scale parallel convolutions (see Section~\ref{section:Parallel}) and channel attention on multi-modal data fusion (see Section~\ref{sec:am}).

Our encoder consists of three down-sampling blocks with $32$, $64$, and $128$ filters, respectively. Only the first encoder block contains our parallel convolution block. The remaining two encoder blocks consist of a squeeze and excitation attention block~\cite{hu2018squeeze} followed by two consecutive convolutions with instance normalization~\cite{ulyanov2016instance}, and LeakyReLU activation with a negative slope of $0.1$. The resulting tensor is passed through a dropout layer at a rate of $0.2$. For each encoder block, an identity skip connection is added to map the encoder blocks onto their corresponding decoder blocks. All convolutions have a kernel size of $3 \times 3 \times 3$, and down-sampling is achieved by $2 \times 2 \times 2$ max-pooling to reduce the spatial resolution of the feature maps.

The decoder takes the feature maps of the encoder and doubles their spatial resolution using 3D trilinear up-sampling. It has three up-sampling blocks, each consisting of two $3 \times 3 \times 3$ convolutions with $128$, $64$, and $32$ filters respectively, followed by instance normalization, and LeakyReLU. A squeeze and excitation attention block has been added between the two convolutions. This is followed by a dropout layer, implemented with a rate of $0.2$ to mitigate overfitting by randomly deactivating a portion of the neural connections during training. After the dropout layer, an additional $3 \times 3 \times 3$ convolutional layer is incorporated. This layer is pivotal for refining the feature representations post-dropout, ensuring the restoration of spatial dimensions, and enhancing the network's ability to learn detailed, spatially coherent features essential for accurate segmentation.

The last two blocks of the encoder and decoder use an L2 regularizer. Finally, a $1 \times 1 \times 1$ convolutional layer with softmax activation is applied to the output of the decoder, which generates a probability map for each voxel indicating its likelihood of belonging to one of the tumor region classes to be segmented.

The LATUP-Net architecture is intentionally designed to minimize the dependence on complex ensembling or additional computational resources. In alignment with best practices outlined in~\cite{isensee2024nnu}, we focus on delivering an efficient, lightweight model that demonstrates its innovations without depending on confounding performance boosters. This approach ensures fair comparisons and highlights the true impact of our architectural choices. By combining proven techniques, such as multi-scale parallel convolutions and lightweight attention mechanisms, LATUP-Net balances the need for high segmentation performance with reduced computational complexity, making it suitable for resource-constrained environments~\cite{bca}.

In line with findings by Rajamani \emph{et al.}~\cite{rajamani2023attention}, integrating an attention block at the network's bottleneck---specifically, an SE block in our case---facilitates the capture of longer-range dependencies within the lowest-resolution activation maps. This adjustment boosts performance with only a slight increase in model complexity.

\begin{table*}[t]
\centering
\caption{LATUP-Net model architecture details.}
\label{tab:model_architecture}
\resizebox{\textwidth}{!}{%
\begin{tabular}{|c|c|c|c|c|c|}
\hline
\textbf{Layer} & \textbf{Input Shape} & \textbf{Layer Type} & \textbf{Stride} & \textbf{Output Shape} & \textbf{Parameters} \\ \hline
Input Layer & (128, 128, 128, 3) & Input & - & (128, 128, 128, 3) & 0 \\ \hline
Conv3D\_1 (enc1\_pc\_embed) & (128, 128, 128, 3) & Conv3D (32 filters) & (1,1,1) & (128, 128, 128, 32) & 2624 \\ \hline
Conv3D\_2 (enc1\_pc\_1\_conv) & (128, 128, 128, 32) & Conv3D (32 filters) & (1,1,1) & (128, 128, 128, 32) & 1056 \\ \hline
Conv3D\_3 (enc1\_pc\_2\_conv) & (128, 128, 128, 32) & Conv3D (32 filters) & (1,1,1) & (128, 128, 128, 32) & 27680 \\ \hline
Conv3D\_4 (enc1\_pc\_3\_conv) & (128, 128, 128, 32) & Conv3D (32 filters) & (1,1,1) & (128, 128, 128, 32) & 128032 \\ \hline
MaxPooling3D\_1 (enc1\_pc\_1\_maxpool) & (128, 128, 128, 32) & MaxPooling3D & (2,2,2) & (64, 64, 64, 32) & 0 \\ \hline
MaxPooling3D\_2 (enc1\_pc\_2\_maxpool) & (128, 128, 128, 32) & MaxPooling3D & (2,2,2) & (64, 64, 64, 32) & 0 \\ \hline
MaxPooling3D\_3 (enc1\_pc\_3\_maxpool) & (128, 128, 128, 32) & MaxPooling3D & (2,2,2) & (64, 64, 64, 32) & 0 \\ \hline
Concatenate (enc1\_pc\_concat) & (64, 64, 64, 32) & Concatenate & - & (64, 64, 64, 96) & 0 \\ \hline
SE Layer\_1 (enc2\_SE\_mult) & (64, 64, 64, 96) & Squeeze and Excitation & - & (64, 64, 64, 96) & 2304 \\ \hline
Conv3D\_5 (enc2\_conv1) & (64, 64, 64, 96) & Conv3D (64 filters) & (1,1,1) & (64, 64, 64, 64) & 165952 \\ \hline
InstanceNorm\_1 (enc2\_instance\_norm) & (64, 64, 64, 64) & Instance Normalization & - & (64, 64, 64, 64) & 128 \\ \hline
Conv3D\_6 (enc2\_conv2) & (64, 64, 64, 64) & Conv3D (64 filters) & (1,1,1) & (64, 64, 64, 64) & 110656 \\ \hline
Dropout\_1 (enc2\_dropout) & (64, 64, 64, 64) & Dropout & - & (64, 64, 64, 64) & 0 \\ \hline
MaxPooling3D\_4 (enc2\_maxpool) & (64, 64, 64, 64) & MaxPooling3D & (2,2,2) & (32, 32, 32, 64) & 0 \\ \hline
SE Layer\_2 (enc3\_SE\_mult) & (32, 32, 32, 64) & Squeeze and Excitation & - & (32, 32, 32, 64) & 1024 \\ \hline
Conv3D\_7 (enc3\_conv1) & (32, 32, 32, 64) & Conv3D (128 filters) & (1,1,1) & (32, 32, 32, 128) & 221312 \\ \hline
InstanceNorm\_2 (enc3\_instance\_norm) & (32, 32, 32, 128) & Instance Normalization & - & (32, 32, 32, 128) & 256 \\ \hline
Conv3D\_8 (enc3\_conv2) & (32, 32, 32, 128) & Conv3D (128 filters) & (1,1,1) & (32, 32, 32, 128) & 442496 \\ \hline
Dropout\_2 (enc3\_dropout) & (32, 32, 32, 128) & Dropout & - & (32, 32, 32, 128) & 0 \\ \hline
MaxPooling3D\_5 (enc3\_maxpool) & (32, 32, 32, 128) & MaxPooling3D & (2,2,2) & (16, 16, 16, 128) & 0 \\ \hline
SE Layer\_3 (bn\_SE\_mult) & (16, 16, 16, 128) & Squeeze and Excitation & - & (16, 16, 16, 128) & 4096 \\ \hline
UpSampling3D\_1 (dec3\_upsample) & (16, 16, 16, 128) & UpSampling3D & (2,2,2) & (32, 32, 32, 128) & 0 \\ \hline
Conv3D\_9 (dec3\_conv1) & (32, 32, 32, 128) & Conv3D (128 filters) & (1,1,1) & (32, 32, 32, 128) & 131200 \\ \hline
InstanceNorm\_3 (dec3\_instance\_norm) & (32, 32, 32, 128) & Instance Normalization & - & (32, 32, 32, 128) & 256 \\ \hline
Concatenate (dec3\_concat) & (32, 32, 32, 128) & Concatenate & - & (32, 32, 32, 256) & 0 \\ \hline
Conv3D\_10 (dec3\_conv2) & (32, 32, 32, 256) & Conv3D (128 filters) & (1,1,1) & (32, 32, 32, 128) & 884864 \\ \hline
UpSampling3D\_2 (dec2\_upsample) & (32, 32, 32, 128) & UpSampling3D & (2,2,2) & (64, 64, 64, 128) & 0 \\ \hline
Conv3D\_11 (dec2\_conv1) & (64, 64, 64, 128) & Conv3D (64 filters) & (1,1,1) & (64, 64, 64, 64) & 65600 \\ \hline
InstanceNorm\_4 (dec2\_instance\_norm) & (64, 64, 64, 64) & Instance Normalization & - & (64, 64, 64, 64) & 128 \\ \hline
Concatenate (dec2\_concat) & (64, 64, 64, 64) & Concatenate & - & (64, 64, 64, 128) & 0 \\ \hline
Conv3D\_12 (dec2\_conv2) & (64, 64, 64, 128) & Conv3D (64 filters) & (1,1,1) & (64, 64, 64, 64) & 221248 \\ \hline
UpSampling3D\_3 (dec1\_upsample) & (64, 64, 64, 64) & UpSampling3D & (2,2,2) & (128, 128, 128, 64) & 0 \\ \hline
Conv3D\_13 (dec1\_conv1) & (128, 128, 128, 64) & Conv3D (32 filters) & (1,1,1) & (128, 128, 128, 32) & 16416 \\ \hline
Concatenate (dec1\_concat) & (128, 128, 128, 32) & Concatenate & - & (128, 128, 128, 64) & 0 \\ \hline
Conv3D\_14 (dec1\_conv2) & (128, 128, 128, 64) & Conv3D (32 filters) & (1,1,1) & (128, 128, 128, 32) & 55328 \\ \hline
Conv3D\_15 (prob\_map) & (128, 128, 128, 32) & Conv3D (4 filters) & (1,1,1) & (128, 128, 128, 4) & 132 \\ \hline
\multicolumn{5}{|r}{\textbf{Total Parameters}} & \multicolumn{1}{|r|}{\textbf{3,069,060}} \\ \hline
\end{tabular}%
}
\end{table*}

\subsection{Parallel Convolutions (PC)}\label{section:Parallel}

CNNs have demonstrated efficacy in feature extraction. However, indiscriminate augmentation of network layers might precipitate overfitting and computational overheads~\cite{chen2014research}. A balance between network depth and width remains paramount. Our strategy employs parallel convolutions with varying kernel sizes, drawing inspiration from the inception model~\cite{szegedy2015going}. This design allows the network to capture features at different scales, yielding a more efficient model with enhanced segmentation performance.

To improve the representation power of the network, which is a key factor in improving its accuracy and reliability, parallel convolutional layers can be added to different encoder and decoder blocks. However, adding parallel convolutions to all blocks may result in overfitting due to the large number of learnable parameters and limited training data. Therefore, it is added only to the first block of the encoder to extract the most fundamental and diverse features from the input data. Parallel convolutions can capture these features at different scales and orientations. This way, we improve representation power while reducing the risk of overfitting.

\begin{figure}
  \centering
  \includegraphics[width=\columnwidth]{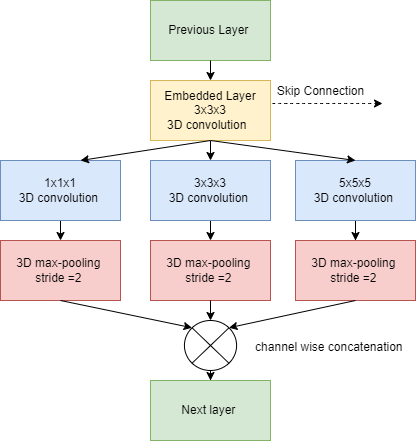}
  \caption{Proposed parallel convolutions.}\label{Figure:parrlele}
\end{figure}

The proposed PC block (see Figure~\ref{Figure:parrlele}) is designed to process the input through a series of convolutional layers with different kernel sizes, each aiming to capture features at various spatial scales. Initially, the input passes through a shared embedded layer of a single $3 \times 3 \times 3$ convolution, which extracts a preliminary set of features from the input data. Following this, the features are processed in parallel through three distinct paths: one continues directly from the initial $1 \times 1 \times 1$ convolution, another passes through an additional $3 \times 3 \times 3$ convolution, and the third through a $5 \times 5 \times 5$ convolution. Convolutional layers with smaller kernel sizes, such as  $1 \times 1 \times 1$ or  $3 \times 3 \times 3$, are adept at detecting local patterns like edges and textures. Layers with larger kernels, like $5 \times 5 \times 5$, are suited for identifying broader spatial patterns and hierarchical structures within the data, thereby providing an extended receptive field. Each path then concludes with a max-pooling operation, reducing the dimensionality and computational load of the subsequent layers. The outputs of these parallel paths are concatenated, combining the multi-scale features into a unified feature map that is richer and more informative than what could be obtained from any single path.

This approach contrasts with the inception block~\cite{szegedy2015going}, which typically includes multiple parallel paths starting from the same input, each with different combinations of convolutions and sometimes pooling operations, without a shared embedded convolution. The inception block aims to capture multi-scale information by applying various-sized convolutions (e.g.,  $1 \times 1 \times 1$,  $3 \times 3 \times 3$,  $5 \times 5 \times 5$) in parallel and then merging their outputs. However, each path in an inception block operates independently of the others, without a shared feature extraction stage. This increases the number of parameters and may lead to redundancy, as each path may learn similar features.

Our proposed PC block contributes to making the model lightweight in several ways. Firstly, the shared embedded layer ensures that all paths operate on a common set of features, reducing redundancy and the need for each path to learn from scratch. This decreases the number of parameters compared to having multiple independent paths, as seen in inception blocks~\cite{szegedy2015going}. Secondly, by limiting each path to a single convolution and a pooling operation after the shared convolution, the model avoids the parameter growth associated with stacking multiple convolutions in each path. This streamlined approach enables efficient multi-scale feature extraction without the complexity and parameter overhead typically associated with more elaborate multi-path designs. Consequently, this design choice balances capturing diverse spatial features and maintaining a compact, efficient model architecture.

\subsection{SE Attention Block}\label{sec:am}

The attention mechanisms explored in this study include Squeeze-\hspace{0pt}and-\hspace{0pt}Excitation (SE)~\cite{hu2018squeeze}, the Convolutional Block Attention Module (CBAM)~\cite{woo2018cbam}, Efficient Channel Attention (ECA)~\cite{wang2020eca}, and Residual Squeeze-\hspace{0pt}and-\hspace{0pt}Excitation (RSE)~\cite{gu2019deep}. We further introduce a modified variant of SE where the fully connected (dense) layers are replaced by a 3D convolutional layer. We also experiment with combined mechanisms such as fusing CBAM and SE. The motivation behind combining CBAM and SE is to leverage the strengths of both. CBAM's ability to focus on pertinent spatial and channel features and SE's capacity to recalibrate channel-wise features may enhance the model's ability to capture complex interdependencies in the data. Another approach is designed to exploit the strengths of convolutions while adhering to the principles of the SE mechanism. Replacing the dense layer in SE with a 3D convolution layer (SE-3D) is aimed at maintaining the spatial information of the input tensor and capturing local spatial correlations, while simultaneously maintaining the ability to recalibrate channel-wise features.

\begin{figure}[t]
  \centering
  \includegraphics[width=0.6\columnwidth, height=8cm]{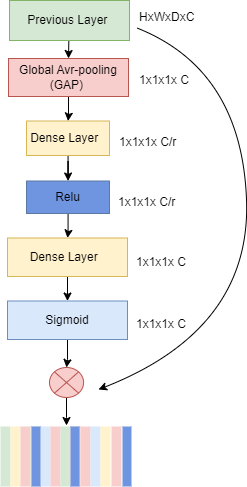}
  \caption{Squeeze and Excitation Block.}\label{Figure:SE}
\end{figure}

Based on rigorous evaluation of attention mechanisms in Section~\ref{section:result}, we incorporate an SE block~\cite{hu2018squeeze} into our final model, which is illustrated in Figure~\ref{Figure:SE}. This block is recognized for its efficiency and lightweight nature. The SE block is composed of two distinct operations: Squeeze and Excitation.
In the squeeze phase, input images of size $H \times W \times D \times C$ are transformed to a $1 \times 1 \times 1 \times C$ format through a Global Average Pooling (GAP) layer, which compresses the spatial resolutions, retaining only channel-centric information for the subsequent excitation operation. The excitation phase employs a series of layers, beginning with a fully connected layer complemented by a reduction factor $r$. This is then subjected to ReLU activation, succeeded by another fully connected layer, culminating in a sigmoid activation to produce the final output of the Excitation operation. A scaling transformation is executed to assimilate the channel-specific data, yielding an output enriched with channel-level information.

\begin{figure*}
  \centering
  \includegraphics[width=\textwidth]{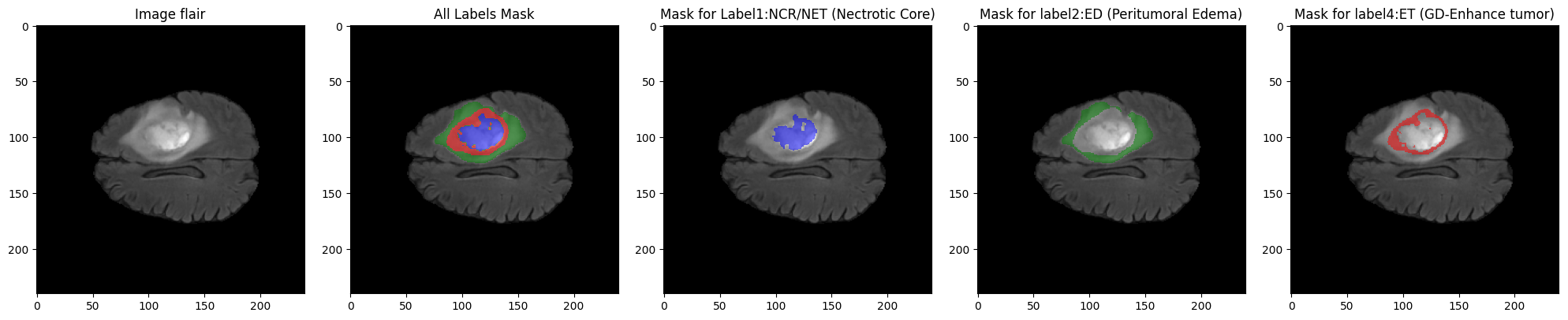}
  \caption{MRI scan of a brain tumor featuring ground truth segmentation masks: blue represents necrotic and non-enhancing tumor areas (NCR/NET, Label~1), green highlights edema regions (ED, Label~2), and red indicates areas of enhancing tumor (ET, Label~4).}\label{Figure:MRI}
\end{figure*}

\section{Data and Implementation Details}\label{sec:details}

\subsection{Data and Pre-processing}\label{sec:data}

The proposed model is trained and validated using the Brain Tumor Segmentation (BraTS) benchmark datasets: BraTS~2020~\cite{menze2014multimodal,bakas2017advancing,bakas2018identifying} and BraTS~2021~\cite{menze2014multimodal}. BraTS~2021, a superset of BraTS~2020, encompasses $1,251$ patients, including high-grade gliomas (HGG) and low-grade gliomas (LGG). The BraTS~2020 dataset contains $369$ patients, of which $76$ have been diagnosed with LGG, with the remainder having HGG.

The BraTS~2020 and BraTS~2021 datasets are used for efficiency and sustainability, with initial testing on the smaller dataset allowing for algorithm refinement and model selection before deployment on the larger dataset. It conserves computational resources and time, and aligns with iterative development best practices, where initial testing on a subset of the data can provide quick feedback, critical for early-stage model tuning and optimization~\cite{raschka2018model}.

Each dataset consists of 3D scans with $155$ individual ``slices'' or images, each having $240 \times 240$ pixels. These scans capture four MRI modalities---T2, T1, T1ce, and FLAIR---crucial for brain tumor segmentation, offering distinct insights. T1-weighted images illustrate anatomical structures, distinguishing between gray and white matter. T2-weighted images aid in visualizing edema, by emphasizing water content. T1ce images, with contrast enhancement, focus on blood vessel imaging, a key component in identifying active tumor growth. Conversely, FLAIR images suppress cerebrospinal fluid signals, illuminating anomalies in intensity and subtle lesions indicative of tumor expansion. Ground truth segmentation masks are meticulously annotated by one to four expert neuroradiologists per case. The scan has been segmented into four primary classes: background (BG, Label~0), necrotic and non-enhancing tumor (NCR/NET, Label~1), edema (ED, Label~2), and enhancing tumor (ET, Label~4) as exemplified in Figure~\ref{Figure:MRI}. Following common practices in the literature, we classify these into three main tumor regions for segmentation: whole tumor (WT) encompassing NCR/NET, ED, and ET (Labels~1,~2,~4), tumor core (TC) consisting of NCR/NET and ET (Labels~1,~4), and the sole ET (Label~4).

The BraTS datasets have been meticulously preprocessed by their developers, including co-registration to a consistent anatomical template, interpolation to a unified resolution ($1~\text{mm}^3$), and skull stripping. However, MRI scan intensity values often display inconsistencies and may fluctuate due to many factors. To mitigate this, we normalize the intensity range to the interval $[0,1]$ using the min-max scaler~\cite{patro2015normalization}. This adjustment not only enhances data consistency but also optimizes it for deep learning algorithms.

We further crop the images to a standard size of $128 \times 128 \times 128$ voxels, centered on the MRI scans. Preliminary tests suggested that including T1 images with T1ce, T2, and FLAIR only marginally improves segmentation results. Since T1ce is essentially a contrast-enhanced derivative of T1, and T1 mainly contributes to identifying a small fraction of the edema, which FLAIR can effectively detect as well, we chose to exclude T1 from our inputs to conserve computational resources.

\subsection{Implementation Details}

We implemented our network via Keras in Tensorflow 2.15. Computations are executed on a single NVIDIA GeForce RTX3060 12~GB GPU, which is considered a relatively low-end consumer card. 
For training, we employed the ADAM optimizer~\cite{kingma2014adam}, setting the learning rate to $1 \times 10^{-4}$. Training proceeds with a batch size of $1$, a choice primarily dictated by GPU memory constraints. We used a constant number of $200$ epochs. To mitigate the risk of overfitting and enhance the model's generalization capabilities, L2 regularization was applied to the convolutional kernel parameters with a factor of $0.02$. This regularization factor was selected based on experimentation during the model selection process, as detailed in the supplementary material.

Leaky ReLU with a leak factor of $\alpha = 0.1$ was used as the activation function for the hidden layers. This value is widely used in deep learning and was chosen based on its effectiveness in similar tasks, such as image segmentation, where it helps regularize the model by allowing gradient flow for negative inputs, contributing to stable training. Studies, including nnU-Net~\cite{isensee2021nnu} and the original leaky ReLU paper~\cite{maas2013rectifier}, 
have shown that $\alpha = 0.1$ works well in practice.

During preliminary experimentation, we evaluated the \texttt{ReduceLROnPlateau} callback for dynamic learning rate adjustments. However, observations indicated a predisposition towards overfitting when it was employed. As such, it was excluded from the final training (see Section~\ref{sec:lr}).

The specific hyper-parameter settings we adopted during model training are detailed in Table~\ref{tbl1}. Detailed results with a discussion are in Section~\ref{section:result} with additional information about the parameter selection process available in the supplementary material. The source code is available at~\cite{bca} with final models and analysis results at~\cite{bca-results}.

\begin{table}
\caption{Hyperparameters for the LATUP-Net model.}\label{tbl1}
\begin{tabular*}{\tblwidth}{LLLL}
\toprule
Hyperparameter & Value\\
\midrule
Input size     & $128 \times 128 \times 128 \times 3$\\
Batch size     & $1$\\
Hidden layer activation & Leaky ReLU($\alpha=0.1$)\\
Optimizer      & ADAM ($\beta_1= 0.9$, $\beta_2=0.999$)\\ 
Learning rate  & $1 \times 10^{-4}$ \\
Number of epochs & $200$\\
Loss function  & Weighted Dice score Loss (see Section~\ref{sec:loss})\\
Dropout        & $0.2$\\
Regularization & L2 (factor $0.02$)\\
Output layer activation & Softmax\\
Output size    & $128 \times 128 \times 128 \times 4$\\
\bottomrule
\end{tabular*}
\end{table}

\subsection{Loss Function}\label{sec:loss}

Loss function selection is a critical factor in contemporary deep-learning network designs, especially in the field of brain tumor segmentation. Recent studies indicate that no single popular loss function consistently offers superior performance across various segmentation tasks~\cite{bakas2017segmentation}.

Compound loss functions, which combine two or more types of loss functions, have emerged as the most robust and competitive in different scenarios~\cite{bakas2017segmentation}. In our experiments, we aim to enhance segmentation performance and address the severe class imbalance in the BraTS datasets by combining Dice loss with Binary Cross Entropy (BCE)~\cite{taghanaki2019combo} and Dice loss with focal loss~\cite{lin99p}. However, based on our experiments, these compounded loss function approaches did not significantly outperform Dice loss alone. Therefore, to boost the segmentation performance and solve the class imbalance problem, the loss function used during the final training process is the Weighted Dice score Loss (WDL). 

The Dice score loss for each class $i$, corresponding to the network output channels BG, NCR/NET, ED, and ET, is
\begin{equation}\label{eq:dscloss}
DSL_{i} = 1 - \frac{2 \mathlarger\sum_{n} \left(y^\text{\vphantom{p}true}_{i,n} \odot y^\text{pred}_{i,n}\right) + \epsilon}{\mathlarger\sum_{n} \left(y^\text{\vphantom{p}true}_{i,n}\right)^2 + \mathlarger\sum_{n} \left(y^\text{pred}_{i,n}\right)^2 + \epsilon}.
\end{equation}
$y^\text{\vphantom{p}true}_i$ and $y^\text{pred}_i$ represent the ground truth and predicted segmentation masks for class $i$, respectively; $n$ iterates over all elements of $y^\text{\vphantom{p}true}_i$ and $y^\text{pred}_i$; $\odot$ signifies point-wise multiplication; and $\epsilon$ is a negligible constant introduced to avoid division by zero. In our experiments, we set $\epsilon = 0.00001$. Note that here the network output masks are the original BG, NCR/NET, ED, and ET regions in the ground truth (see Section~\ref{sec:data}).

The WDL weights $w_i$ for class $i$ are computed according to the ENet paper~\cite{paszke2016enet},
\begin{equation}
w_i = \frac{1}{\log(C + \frac{c_i}{T})}
\end{equation}
where $C = 1.02$, $c_i$ is the voxel count for class $i$, and $T$ is the total count of voxels across all classes. This formula ensures that classes with fewer voxels receive higher weights to balance the loss during the training process. Here we use the WT, TC and ET regions instead of the individual output channels above to compute the weights, and we get $w_{\text{WT}} = 1.64$, $w_{\text{TC}} = 2.55$, and $w_{\text{ET}} = 3.40$.

Overall this gives our Weighted Dice score Loss (WDL),
\begin{equation}
\begin{split}
\text{WDL} = & \; w_{\text{WT}} \cdot (DSL_{\text{NCR/NET}} + DSL_{\text{ED}} + DSL_{\text{ET}}) \\
             & + w_{\text{TC}} \cdot (DSL_{\text{NCR/NET}} + DSL_{\text{ET}}) \\
             & + w_{\text{ET}} \cdot DSL_{\text{ET}},
\end{split}
\end{equation}
which is equivalent to
\begin{equation}
\begin{split}
\text{WDL} = & \; (w_{\text{WT}} + w_{\text{TC}} + w_{\text{ET}}) \cdot DSL_{\text{ET}} \\
             & + (w_{\text{WT}} + w_{\text{TC}}) \cdot DSL_{\text{NCR/NET}} \\
             & + w_{\text{WT}} \cdot DSL_{\text{ED}}.
\end{split}
\end{equation}
In this expression, the dice score loss (DSL) for each class $i \in \{\text{NCR/NET}, \text{ED}, \text{ET}\}$ is computed separately and weighted according to the importance of the corresponding tumor region (WT, TC, and ET) and then summed up to compute the total weighted Dice score loss for the segmentation task. This loss function links the clinical relevance of each tumor region with the network's output channels, ensuring that the segmentation process prioritizes the most clinically significant areas which is crucial for achieving optimal segmentation performance. Using the ENet weights helps in addressing the class imbalance by assigning higher weights to smaller but more important tumor regions.

It is also important to note that while the network output channels include the background class (BG), it is not included in the loss and weights calculation. We found its presence improves the segmentation performance of the model. We also explored the use of different output channels, weights, and manual adjustment of the weights. However, it did not yield satisfactory results.

\subsection{Evaluation Metrics}

We measure the effectiveness of the proposed model using the Dice similarity coefficient (DSC), and the $95^\mathit{th}$ percentile Hausdorff distance (HD95). DSC and HD95 are widely accepted as the primary performance evaluation metrics in image segmentation tasks. The DSC quantifies the spatial overlap between the ground truth and the predicted segmentation region. HD95 calculates the $95^\mathit{th}$ percentile of the distances between the points in the ground truth and the predicted set. This is akin to the conventional symmetric Hausdorff distance but reduces the impact of outliers by focusing on the $95^\mathit{th}$ percentile. HD95 in particular indicates the accuracy of boundary prediction, revealing the model's precision in delineating the tumor margins.

The metrics are defined as
\begin{gather}
\mathrm{DSC} =\frac{2 |P \cap T|}{|P| + |T|},\\
\begin{multlined}
\text{HD95} = \max\left(\max_{p \in P_{95\%}} \min_{t \in T} \lVert p - t \rVert,\right.\\
                  \left.\max_{t \in T_{95\%}} \min_{p \in P} \lVert t - p \rVert\right),
\end{multlined}
\end{gather}
where $p$ and $t$ are voxel coordinates for the predicted and ground truth regions respectively; $P$ is the predicted region, and $T$ is the ground truth region. The corresponding $95^\mathit{th}$ percentile regions are represented by $P_{95\%}$ and $T_{95\%}$.

We employ different metrics for evaluating our model based on the data partitioning approach used. For model selection (see Section~\ref{sec:performance}, and Section~\ref{sec:att-cmp}), we employ an $80$/$20$ training-testing holdout split, and we report the mean and standard deviation of the per-sample (per-patient) metrics to gain insights into the model's performance under controlled conditions. Once the optimal model (LATUP-Net) is determined, and for comparison to the state-of-the-arts models (see Section~\ref{section:comparasion}), five-fold cross-validation is applied to consider any data dependencies and ensure a comprehensive evaluation of the model's performance across varied data scenarios. During cross-validation, we calculate the mean and standard deviation of the mean DSC and HD95 across all folds to assess the model's robustness and general performance. 

\section{Experimental Results and Discussion}\label{section:result}

This section covers four main analyses. Firstly, we analyze the performance of some variants of our architecture and the influence of learning rate optimization on the segmentation performance. Then, we investigate which attention mechanism gives the best performance. We also evaluate the effectiveness of the attention mechanism. Finally, we compare the performance of our LATUP-Net to the state-of-the-art on BraTS~2020 and~2021.

\subsection{Overall Performance Analysis} \label{sec:performance}

To find the best variant of our proposed model and training, we compare the following architectures:
\begin{description}
\item[U-Net:] The baseline U-Net model trained using Dice loss.
\item[Inception:] Modified U-Net model with inception module trained using Dice loss.
\item[PC:] Modified U-Net model with parallel convolutions trained using Dice loss.
\item[PC + SE:] Modified U-Net model with parallel convolutions and channel attention trained using Dice loss.
\item[PC + WDL:] Modified U-Net model with parallel convolutions trained using weighted Dice score Loss.
\item[PC + SE + WDL:] Modified U-Net model with parallel convolutions, and attention trained using weighted Dice score Loss.
\end{description}

The models are initially selected based on a single $80$/$20$ split using the same training hyperparameters (see Table~\ref{tbl1}) on the BraTS~2020 dataset. Table~\ref{table2} shows the segmentation results of the test set from these training runs. The performance is assessed by employing two key metrics: per-sample DSC and HD95.

\begin{table*}[t]
\caption{Comparative analysis of model architectures for brain tumor segmentation using the BraTS~2020 dataset: This table illustrates the mean and standard deviation (indicated by $\pm$) for the per-sample Dice similarity coefficient (DSC) and the $95^\mathit{th}$ percentile Hausdorff distance (HD95). Results are segmented into whole tumor (WT), tumor core (TC), and enhancing tumor (ET) categories, based on an $80$/$20$ train/test set split.}\label{table2}
\centering
\SetTblrInner{rowsep=3pt}
\resizebox{\textwidth}{!}{%
\begin{tblr}{
  colspec={Q[2.5cm,valign=m]lllllllll},
  cell{1-2}{1-8}={gray!20},
  cell{1}{1}={r=2}{l},
  cell{1}{2}={c=3}{l},
  cell{1}{5}={c=3}{l},
  cell{1}{8}={r=2}{l}
}
\hline
Model           & DSC ($\%$) &&& HD95 (mm) &&& \\\cline{2-9}
                & WT & TC & ET & WT & TC & ET &\\\hline
U-Net           & $83.22\pm9.29$  & $77.13\pm18.34$ & $60.65\pm27.44$ & $18.50\pm22.17$ & $15.38\pm22.91$ & $19.34\pm30.24$ \\
Inception       & $87.53\pm7.49$  & $80.91\pm18.64$ & $69.28\pm29.05$ & $10.87\pm15.72$ & $6.58\pm7.78$   & $14.94\pm29.06$ \\
PC              & $88.13\pm7.14$  & $84.19\pm16.74$ & $70.23\pm28.40$ & $4.99\pm4.00$   & $5.63\pm6.71$   & $12.86\pm26.39$ & \\
PC + SE         & $88.52\pm7.10$  & $83.26\pm17.18$ & $71.86\pm27.02$ & $5.98\pm7.88$   & $5.51\pm5.20$   & $12.96\pm26.55$ \\
PC + WDL        & $\mathbf{89.58\pm5.70}$ & $\mathbf{85.35\pm15.49}$ & $73.21\pm27.18$ & $\mathbf{4.78\pm4.41}$  & $5.25\pm6.38$  & $11.65\pm26.33$  \\
PC + SE + WDL   & $88.72\pm6.33$ & $84.71\pm15.65$ & $\mathbf{74.49\pm25.98}$ & $5.76\pm4.41$  & $\mathbf{5.15\pm6.28}$  & $\mathbf{10.65\pm25.33}$ \\\hline
\end{tblr}}
\end{table*}

In the Inception model, we replace the first U-Net block with an inception module. This modification improves the segmentation results compared to the U-Net, with DSC improvements of $4.31$, $3.78$, and $8.63$ for whole tumor (WT), tumor core (TC), and enhancing tumor (ET), while reducing the HD95 by $7.63$, $8.8$, and $4.44$ for WT, TC, and ET, respectively. However, despite these performance gains, the Inception model significantly increases memory usage during training, likely due to the module's more complex structure, which combines multiple convolutional and pooling operations.

The PC model, using parallel convolutions, demonstrates superior efficiency compared to both U-Net and Inception. It achieves faster convergence during training and reduces the need for computational resources. Additionally, PC provides substantial improvements in segmentation performance compared to U-Net, with DSC gains of $4.91$, $7.06$, and $9.58$ and HD95 reductions of $13.51$, $9.75$, and $6.48$ for WT, TC, and ET, respectively.

When comparing our proposed parallel convolutions (PC) with the Inception module, PC offers a strategic advantage. By replacing the conventional $1\times1\times1\times1$ convolution with a $3\times3\times3\times3$ convolution, PC achieves a more spatially compact and contextually rich representation, while effectively reducing memory consumption. In contrast, the Inception module with its complex configuration faced a surge in model memory usage during training. Furthermore, our design's judicious positioning of pooling operations optimally condenses feature map dimensions, ensuring efficient memory usage without compromising capturing features. The PC model outperforms the inception model with $0.6$, $3.28$, and $0.95$ DSC increment and $5.88$, $0.95$, and $2.08$ HD95 decrement for WT, TC, and ET, respectively, which proves PC's ability to segment difficult tumor regions such as TC.

The addition of SE attention to the PC model (PC + SE) further enhances segmentation performance, particularly for smaller tumor regions, as evidenced by DSC improvements of $0.39$ for WT and $1.63$ for ET. However, a slight reduction in TC performance ($0.93$ decrease in DSC) suggests that while attention mechanisms provide benefits in some areas, they may introduce trade-offs for certain tumor regions. Several lightweight attention modules are compared in Section~\ref{sec:att-cmp}.

Notably, our investigations reveal a significant performance enhancement upon employing the WDL with the PC model. Specifically, the DSC for WT, TC, and ET increased by $1.45$, $1.16$, and $2.98$ respectively and HD95 reduced by $0.21$, $0.38$, and $1.21$ respectively.

This improvement underscores the utility of the WDL in case of a segmentation region size imbalance. However, we notice that when adding the attention mechanism, the result of WT and TC decreased slightly in both DCS and HD95, which leads us to check whether attention is needed. Nonetheless, we chose to include SE because there is a 1.28\% improvement in the ET segmentation result and enhancements in HD95 for TC and ET. This is further investigated in Section~\ref{sec:att-need}.

\begin{figure*}[t]
\centering
\hspace*{0pt}\hfill 
\parbox{.13\textwidth}{\centering Flair}\hfill 
\parbox{.13\textwidth}{\centering Ground Truth}\hfill 
\parbox{.13\textwidth}{\centering U-Net}\hfill 
\parbox{.13\textwidth}{\centering Inception}\hfill 
\parbox{.13\textwidth}{\centering PC}\hfill 
\parbox{.13\textwidth}{\centering PC+SE}\hfill 
\parbox{.13\textwidth}{\centering PC+SE+WDL }\hfill\\
\resizebox{\textwidth}{!}{%
\begin{tabular}{ccccccc}
\includegraphics[width=1\textwidth]{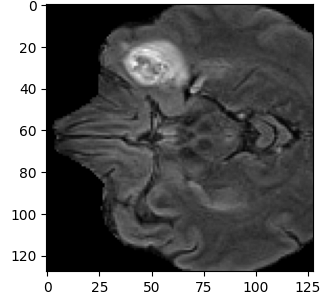} & 
\includegraphics[width=1\textwidth]{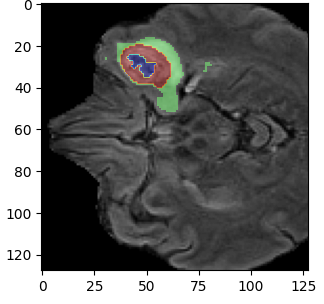} & 
\includegraphics[width=1\textwidth]{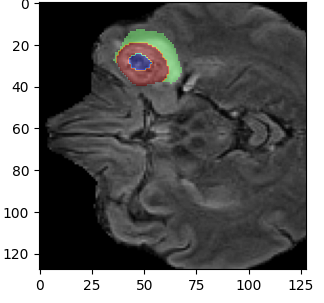} & 
\includegraphics[width=1\textwidth]{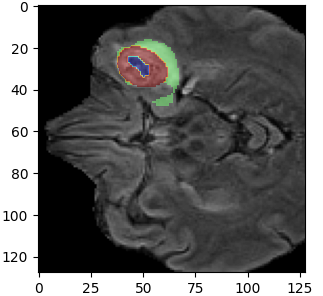} & 
\includegraphics[width=1\textwidth]{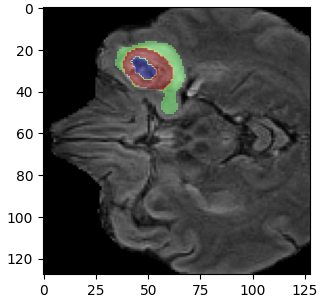} &
\includegraphics[width=1\textwidth]{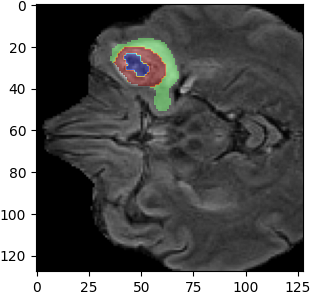} & 
\includegraphics[width=1\textwidth]{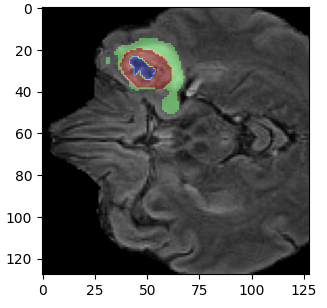} \\
\includegraphics[width=1\textwidth]{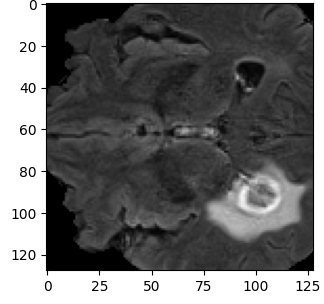} & 
\includegraphics[width=1\textwidth]{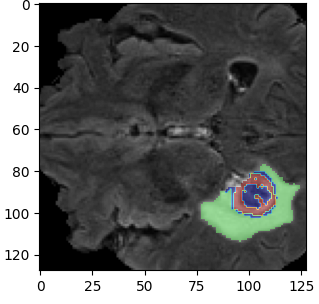} & 
\includegraphics[width=1\textwidth]{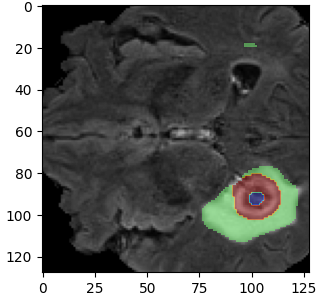} & 
\includegraphics[width=1\textwidth]{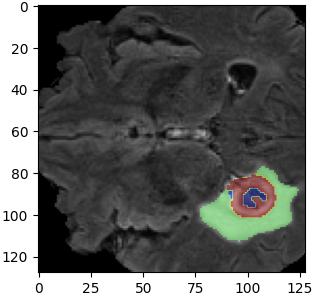} & 
\includegraphics[width=1\textwidth]{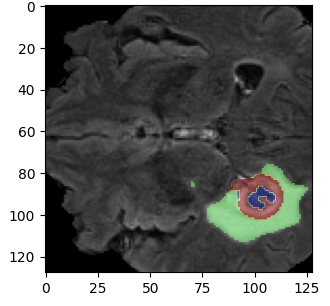} &
\includegraphics[width=1\textwidth]{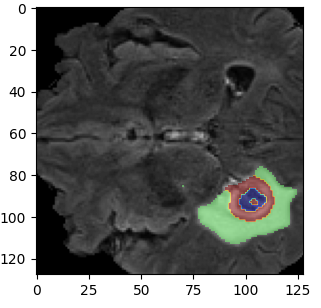} & 
\includegraphics[width=1\textwidth]{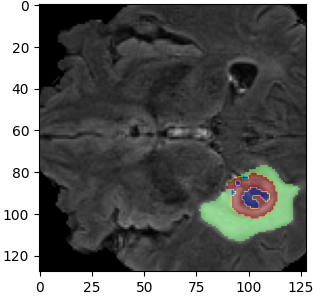} \\
\includegraphics[width=1\textwidth]{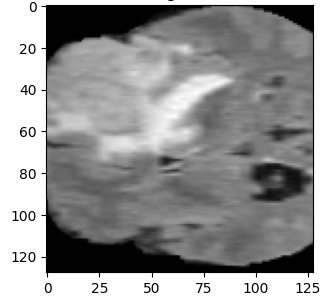} & 
\includegraphics[width=1\textwidth]{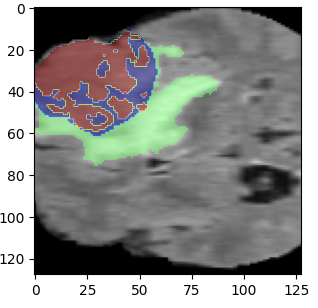} & 
\includegraphics[width=1\textwidth]{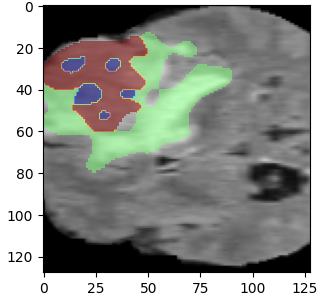} & 
\includegraphics[width=1\textwidth]{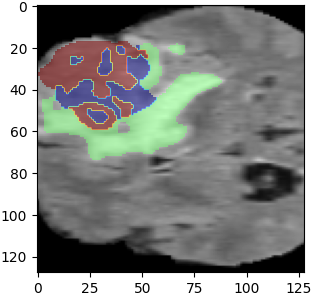} & 
\includegraphics[width=1\textwidth]{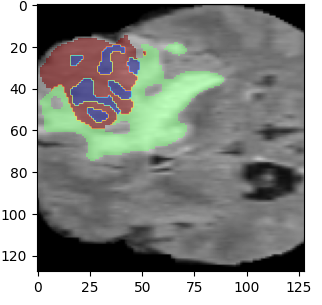} &
\includegraphics[width=1\textwidth]{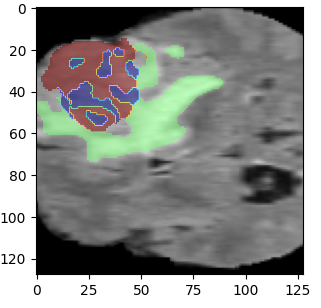} & 
\includegraphics[width=1\textwidth]{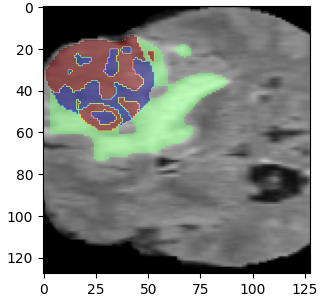} \\
\end{tabular}}
\caption{Segmentation results from the test set that are typical of those produced by the various networks. The results for a single patient from each network are shown in each row. The enhancing tumor is depicted in blue, the necrotic and non-enhancing tumor in red, and the edema in green (after extracting the distinct regions from the partially overlapping segmentation results).}\label{Figure:Picture2}
\end{figure*}

Figure~\ref{Figure:Picture2} depicts qualitative comparisons between the various networks in segmenting the distinct tumor regions. The qualitative results demonstrate that our LATUP-Net which consists of PC and SE trained with WDL, outperformed all other models, consistent with the quantitative results in Table~\ref{table2}.

\subsubsection{The Influence of Learning Rate Decay on the Segmentation Performance}\label{sec:lr}

\begin{figure}
  \centering
  \includegraphics[width=.49\columnwidth]{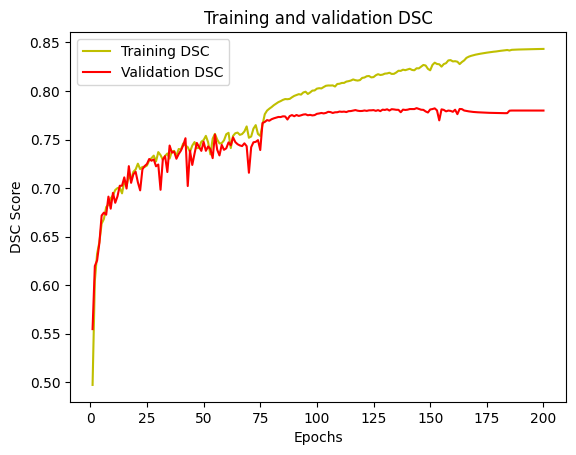}\hfill
  \includegraphics[width=.49\columnwidth]{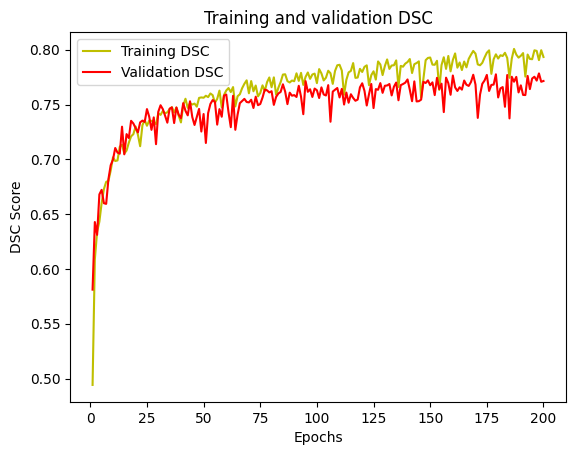}\\
  \hspace*{0pt}\hfill (a) \hfill\hfill (b) \hfill\hspace*{0pt}
  \caption{Modified U-Net with PC: Dice score curves for all tumor regions (a) with and (b) without learning rate decay.}\label{Figure:pc}
\end{figure}

Training neural networks necessitates careful control of convergence and prevention of overfitting. Traditional models typically utilize learning rate schedulers alongside stochastic gradient descent. However, with the introduction of advanced optimizers such as ADAM, which integrates momentum and regularization, there has been a shift to strategies like \texttt{ReduceLROnPlateau}. This approach adjusts the learning rate in response to training plateaus by monitoring validation loss, akin to early stopping criteria~\cite{9861045}. Despite the recognized benefits of \texttt{ReduceLROnPlateau} in contemporary optimization scenarios, our experimentation yielded nuanced insights. While training models with and without learning rate decay, a small performance improvement was observed, but the associated learning curves exhibit clear signs of overfitting (see Figure~\ref{Figure:pc}(a)). In contrast, models trained without learning rate decay did not display such overfitting tendencies (see Figure~\ref{Figure:pc}(b)). These observations are likely attributed to our model's already minimal learning rate. Consequently, in our final model configuration, we opted to forgo the learning rate decay to circumvent the observed overfitting tendencies.

\subsection{Comparison of Different Attention Mechanisms}\label{sec:att-cmp}

In our approach so far, we have used the SE attention mechanism. In this section, we explore whether alternative attention mechanisms (see Section~\ref{sec:ATT}) may yield improved performance within the confines of our model. For a uniform evaluation, we incorporate all attention mechanisms after the parallel convolutions (PC). It is also important to note that the model is trained using Dice loss and the results presented here are derived from configurations that do not include the weighted Dice score loss (WDL). This choice was made to isolate the impact of the attention mechanisms themselves, allowing for a clear comparison of their effectiveness in the absence of the WDL's influence.

Table~\ref{table3} summarizes our findings, detailing the performance metrics of residual SE, CBAM, ECA, and their combinations. While all mechanisms enhance the model's accuracy, SE stands out as the most effective.

CBAM, despite its comprehensive design, might be limited by its complexity, which could result in computational overhead, potentially affecting efficiency, especially in real-time scenarios. On the other hand, while ECA is designed to be lightweight and computationally efficient, it may not capture the intricate inter-channel dependencies as effectively as SE. This trade-off between computational complexity and attention performance suggests that CBAM and ECA may be better suited for specific tasks or resource-constrained environments where accuracy can be sacrificed for speed or memory efficiency.

The SE mechanism's advantage lies in its ability to efficiently highlight enhanced tumors, which is crucial for tasks involving small yet important regions like the tumor core (TC) and enhancing tumor (ET). Additionally, the reduced number of parameters in SE makes it particularly advantageous for large-scale medical image segmentation tasks, where model size and efficiency are important factors. The residual SE mechanism further boosts performance, particularly in terms of Dice similarity coefficient (DSC), as shown by its slightly higher values across all tumor regions.

Analyzing the standard deviation values, the variance in DSC scores appears minimal, which suggests that the choice of attention mechanism might not drastically affect the accuracy. However, in practice, a more streamlined model with fewer parameters (such as PC+SE) may offer competitive or even superior performance due to its balance of effectiveness and computational efficiency.

In contrast, the HD95 metric highlights more prominent differences across the methods, suggesting that attention mechanisms have a varying impact on capturing spatial discrepancies in segmentation, such as outliers or irregularly shaped tumors. For example, CBAM exhibits lower HD95 values for the whole tumor (WT), indicating that its spatial attention might better capture large-scale spatial variations, while SE-based models perform better on smaller, enhancing tumors (ET), where channel-wise attention is more effective at capturing fine-grained details.

Given the observed performance trade-offs, the choice of attention mechanism may depend on the specific task requirements, including segmentation accuracy, computational resources, and the need to balance spatial and channel-wise attention.

\begin{table*}
\caption{Efficiency analysis of various attention mechanisms for brain tumor segmentation on the BraTS~2020 test set. This table illustrates the mean and standard deviation (indicated by $\pm$) for the per-sample Dice similarity coefficient (DSC) and the $95^\mathit{th}$ percentile Hausdorff distance (HD95). Results are segmented into whole tumor (WT), tumor core (TC), and enhancing tumor (ET) categories, based on an $80$/$20$ train/test set split. Additionally, the table includes the number of trainable parameters for each model.}\label{table3}
\centering
\SetTblrInner{rowsep=3pt}
\resizebox{\textwidth}{!}{
\begin{tblr}{
  colspec={Q[2cm,valign=m]lllllllll},
  cell{1-2}{1-8}={gray!20},
  cell{1}{1}={r=2}{l},
  cell{1}{2}={c=3}{l},
  cell{1}{5}={c=3}{l},
  cell{1}{8}={r=2}{l}
}
\hline
Method             & DSC ($\%$) &&& HD95 (mm) &&& \\\cline{2-9}
                   & WT & TC & ET & WT & TC & ET  \\\hline
PC+SE              & $88.52\pm7.10$ & $83.26\pm17.18$ & $\mathbf{71.86\pm27.02}$ & $5.98\pm7.38$ & $\mathbf{5.51\pm5.20}$  &$\mathbf{12.96\pm26.55}$ \\
PC+CBAM            & $89.38\pm6.53$ & $\mathbf{84.36\pm15.94}$& $70.01\pm29.39$ & $\mathbf{4.78\pm4.19}$ & $5.51\pm5.65$ & $14.12\pm28.27$ \\
PC+CBAM+SE         & $88.91\pm6.31$ & $83.87\pm15.63$ & $70.25\pm28.33$ & $5.24\pm5.07$  & $5.58\pm5.60$  & $12.92\pm26.42$ \\
PC+ (SE-3D) & $89.05\pm6.56$ & $83.91\pm16.97$ & $69.83\pm29.63$ & $5.38\pm6.89$  & $5.72\pm7.37$  & $13.75\pm28.14$  \\
PC+Residual SE     & $\mathbf{89.60\pm6.50}$ & $84.06\pm17.38$ & $70.79\pm29.42$ & $5.40\pm7.46$  & $5.93\pm8.16$  & $13.19\pm26.34$  \\
PC+ECA             & $84.47\pm6.97$ & $80.12\pm18.21$ & $63.19\pm28.15$ & $6.81\pm4.01$  & $7.08\pm7.05$ & $15.03\pm28.13$ \\\hline
\end{tblr}}
\end{table*}

\subsection{Lightweight Model Validation}\label{sec:lightweight}

To demonstrate the lightweight nature of LATUP-Net, Table~\ref{table55} presents a comparison with other models on the floating-point operations (FLOPs) required to evaluate the model once, the number of parameters, and the inference times. Fewer FLOPs indicates a more efficient training process, as less computational power is required. Similarly, a lower number of trainable parameters not only reduces training time but also minimizes memory usage during deployment. Additionally, shorter inference times further highlight the lightweight characteristics of the model. It is important to note that there are various methods for estimating FLOPs. In our study, we used the TensorFlow profiler~\cite{tensorflowprofiler} function from the TensorFlow package, ensuring consistency and accuracy.

The baseline 3D U-Net, with $5.65$ million parameters and $23.8$~GFLOPs, is one of the more complex models, particularly in terms of parameters, reflecting its deeper and wider architecture. In contrast, Inception has fewer parameters ($3.65$ million) but a much higher GFLOP count ($74.98$), suggesting that while it has fewer parameters, its increased computational demand arises from its more intricate layer configurations.

Compared to Inception, LATUP-Net (PC + SE) achieves a more balanced and efficient design, with $15.79$~GFLOPs, $3.07$ million parameters, and an inference time of 168 ms, measured using an NVIDIA RTX 3060 GPU. It is important to note that inference and training times are highly dependent on the hardware used, which is why these times are not compared with other state-of-the-art models in this study. Despite this, LATUP-Net is far more efficient computationally. The PC-based variants (PC, PC + SE, and PC + Residual SE) maintain a similar parameter range ($3.06$ -- $3.07$ million) and low GFLOP counts ($15.08$ -- $16.08$), highlighting their efficiency. While models like PC + CBAM and PC + ECA slightly increase GFLOPs, they still maintain efficient inference times. Overall, LATUP-Net provides significant advantages over Inception by delivering better performance with fewer computations, making it a more efficient model, especially in terms of computational complexity and speed.

\begin{table}[]
\caption{Efficiency comparison of different models based on FLOPs (GigaFLOPs), parameter count (millions), and inference time (milliseconds) measured on an NVIDIA RTX 3060 12 GB GPU.}\label{table55}
\resizebox{\columnwidth}{!}{%
\begin{tblr}{
  colspec={lccc},
  cell{1}{1-4}={gray!20}
}
\hline
Models    & Parameters (M) & FLOPs (G) & Inference Time (ms)\\\hline
U-Net     & $5.65$         & $23.8$    & $230$              \\ 
Inception & $3.65$         & $74.98$   & $333$              \\ 
PC        & $\mathbf{3.06}$& $\mathbf{15.08}$ & $227$       \\ 
PC + SE   & $3.07$         & $15.79$   & $\mathbf{168}$     \\
PC + CBAM & $3.45$         & $17.35$   & $247$              \\ 
PC + CBAM + SE & $3.26$    & $16.08$   & $240$              \\ 
PC + (SE-3D) & $3.07$      & $15.79$   & $234$              \\ 
PC + Residual SE & $3.07$  & $15.80$   & $238$              \\ 
PC + ECA  & $3.10$         & $15.80$   & $212$              \\\hline 
\end{tblr}}

\end{table}

\subsection{Evaluating the Effectiveness of the Attention Mechanism}\label{sec:att-need}

To understand the impact of the integrated attention mechanism and to determine if it operates as intended in our proposed architecture, we conduct two primary experiments: gradient-weighted class activation mapping (Grad-CAM) visualizations~\cite{selvaraju2017grad} and confusion matrix analysis.

\subsubsection{Visual Interpretation using Grad-CAM}\label{sec:gc}

In investigating the efficacy of attention mechanisms in brain tumor detection, the question arises whether the local context is sufficient for accurate segmentation. Traditional convolutions inherently capture the local context without necessitating attention. Given the qualitative nature of our experiments and the challenges of visualizing every data point, it is crucial to strategically select representative samples for thorough analysis. To interpret and delve deeper into the model's behavior, Grad-CAM is applied to three distinct samples that are the same for both models, with attention (LATUP-Net) and without attention (PC + WDL). These samples are selected based on their loss value with some representative slice from the test set and represent the best-performing, the median, and the worst-performing w.r.t. the WDL for comparative analysis.

\begin{figure*}[t]
\centering
\renewcommand{\arraystretch}{0.2} 
\centerline{\small Best Predicted Sample}
\includegraphics[width=1\textwidth]{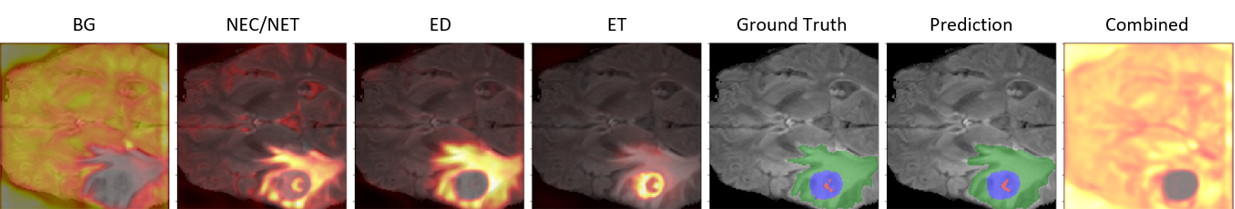} \\
\includegraphics[width=1\textwidth]{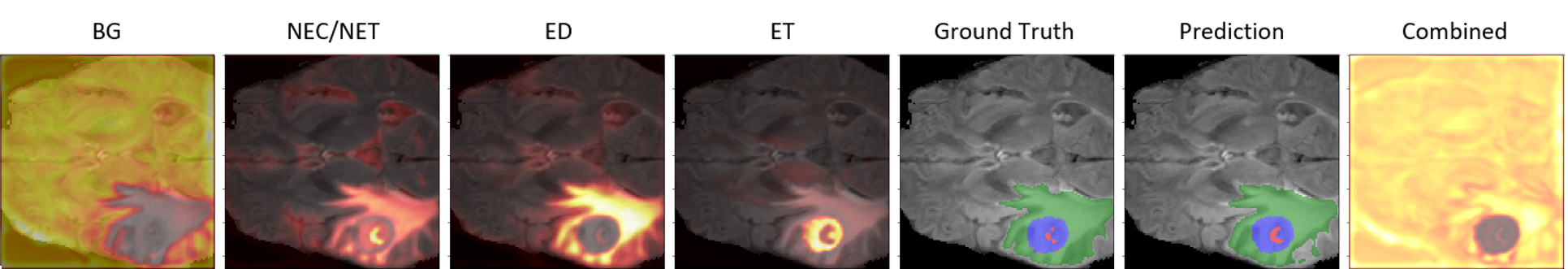} \\
\centerline{\small Median Predicted Sample}
\includegraphics[width=1\textwidth]{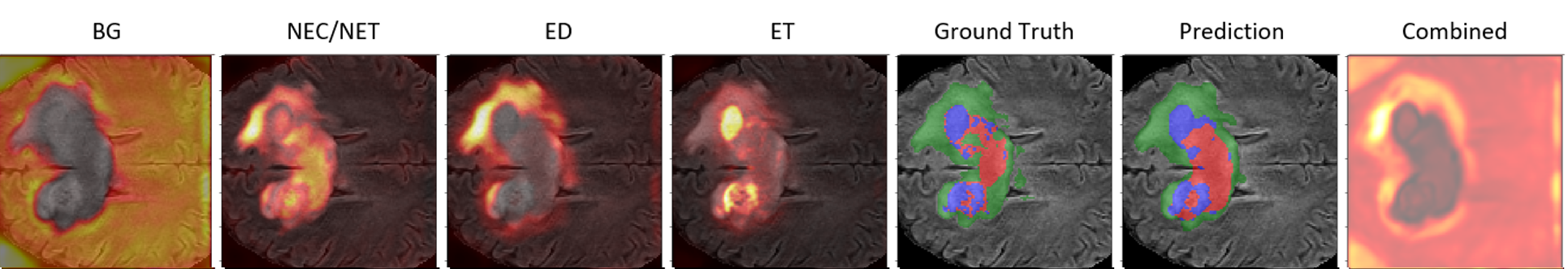} \\
\includegraphics[width=1\textwidth]{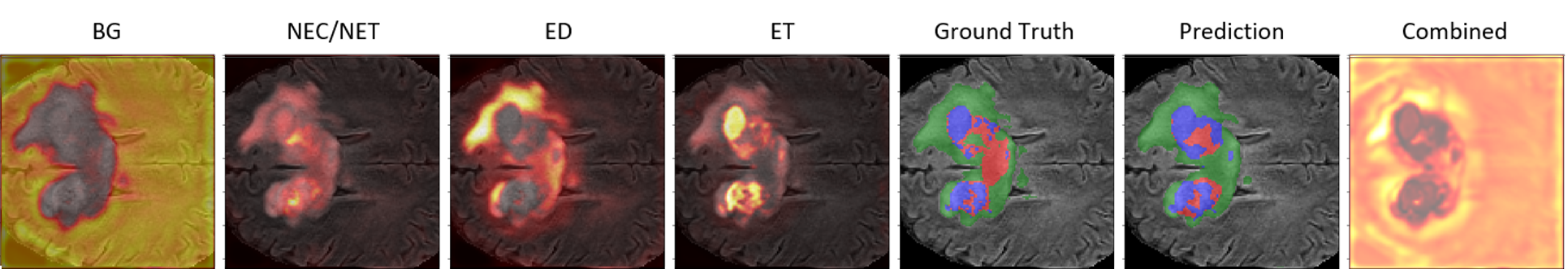} \\
\centerline{\small Worst Predicted Sample}
\includegraphics[width=1\textwidth]{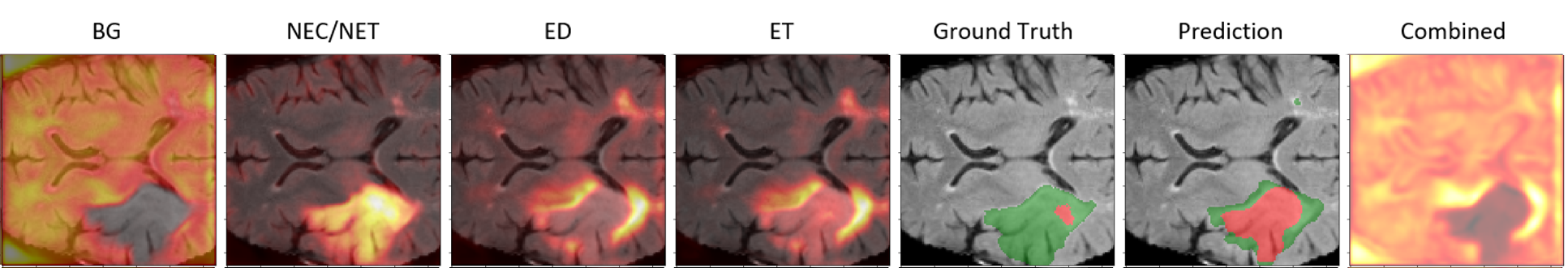} \\
\includegraphics[width=1\textwidth]{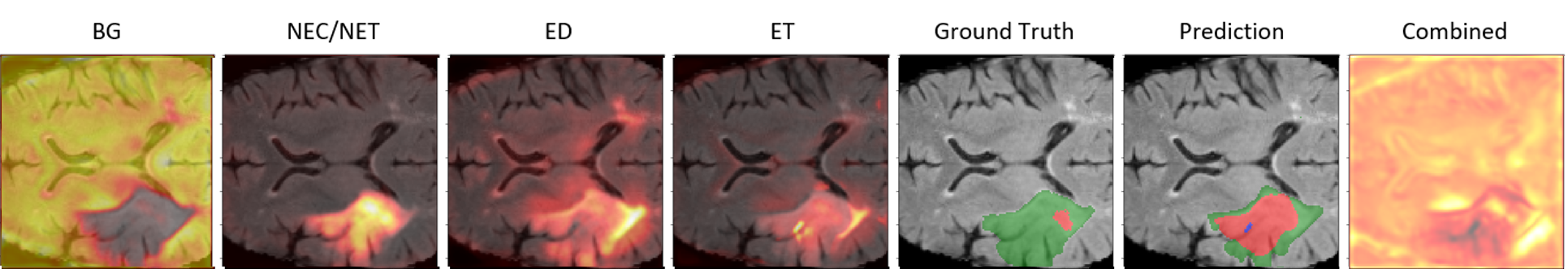}

\caption{Visual interpretations of model predictions using Grad-CAM: For each of the three samples selected from the test set, representing the best, median, and worst predicted cases, both the standard and attention-enhanced models. From left to right in each row we show the GradCAM for the BG, NEC/NET, ED, ET output channel, the ground truth and the prediction, and finally the combined GradCAM for all output channels. The first row visualizes the standard model (PC + WDL), and the second row visualizes the attention-enhanced model (LATUP-Net) per case.}\label{Figure:Picture3}
\end{figure*}

Grad-CAM~\cite{selvaraju2017grad} was applied to the layer preceding the softmax activation used for generating the predictions. This layer was particularly selected because it directly contributes to the decision-making process, offering a transparent view into how the model weighs different regions in its segmentation task. Grad-CAM visualizations, as shown in Figure~\ref{Figure:Picture3}, demonstrate that both models focus primarily on tumor areas. While the attention mechanism intensifies this focus, it does not always lead to the most accurate predictions. In particular, the necrotic region tends to be overemphasized, leading to misclassifications, as necrotic and enhancing regions often share similar textures.

This observation suggests that while attention appears to work as designed by emphasizing certain regions, it may not contribute to improved performance due to its reliance on local features. Grad-CAM further reveals that the segmentation task is heavily driven by these local features, like texture and boundaries, which attention does not necessarily enhance in a meaningful way. Additionally, because Grad-CAM tends to focus on the regions directly involved in segmentation, it does not provide insights into global relationships, such as left-right symmetry of the brain, which the attention mechanism might have been expected to capture.

Overall, the results indicate that the attention mechanism may have limited impact on segmentation performance, as the task is largely dependent on local feature extraction, and the attention mechanism does not provide significant additional context in this particular use case.

\subsubsection{Confusion Matrix Analysis and Implications}

\begin{figure}[t]
  \centering
  \includegraphics[width=\columnwidth]{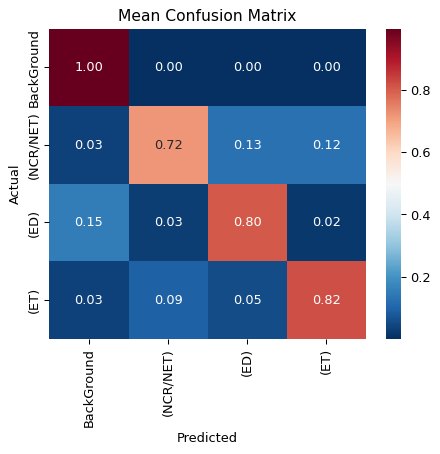}
  \caption{Confusion matrix illustrating the pixel-wise misclassification rates between different tumor regions, normalized by the number of actual class instances in each row and predicted class. These values are averaged over all samples in the dataset to provide a comprehensive overview of the LATUP-Net model's performance.}\label{Figure:ConfusionMatrix}
\end{figure}

\begin{figure}
  \centering
  \includegraphics[width=\columnwidth]{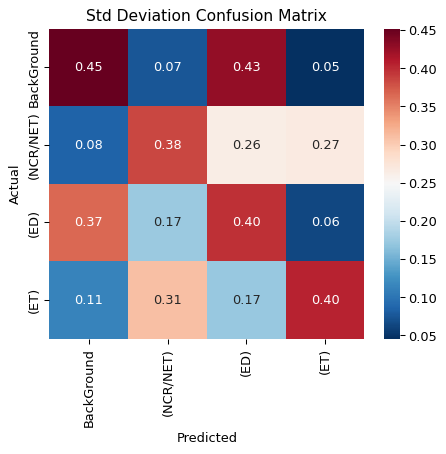}
  \caption{Confusion matrix of the standard deviation of misclassification rates for each tumor region from Figure~\ref{Figure:ConfusionMatrix} of the LATUP-Net model.}
  \label{Figure:StdDeviation}
\end{figure}

To provide quantitative evidence supporting the findings from Grad-CAM visualizations, we perform a confusion matrix analysis, as shown in Figure~\ref{Figure:ConfusionMatrix}. This matrix details how pixels from one class are misclassified as those from another, offering insights into the LATUP-Net model behavior. It is important to note that there were patients with no enhanced tumor (ET) class in the ground truth. These cases have not been considered for the confusion matrix, as including them would distort the percentages. A primary observation is the misclassification between the necrotic and edema regions, a claim also derived from the qualitative visuals (see Section~\ref{sec:gc}). The model perceives textural similarities between these areas, further complicated by nested tumor structures. Additionally, the necrotic region is notably vulnerable to misclassification, perhaps due to its texture and position within the tumor's structure. When examining the standard deviation in Figure~\ref{Figure:StdDeviation}, combined with the average misclassification percentages, the classification performance variability is clear. This variability might spotlight outliers or deviations in model predictions, with classes like the necrotic region and enhancing region showing significant misclassification variability. The variations imply that certain samples heavily influence averages. In conclusion, the attention mechanism, while offering nuanced insights, can overly prioritize local features, causing the model to overlook important topological relationships recognized by human experts. A balance between local and global contexts seems important.

\subsection{Comparison with State-of-the-Art Models}\label{section:comparasion}

\begin{figure*}
  \centering
  \includegraphics[width=.9\textwidth]{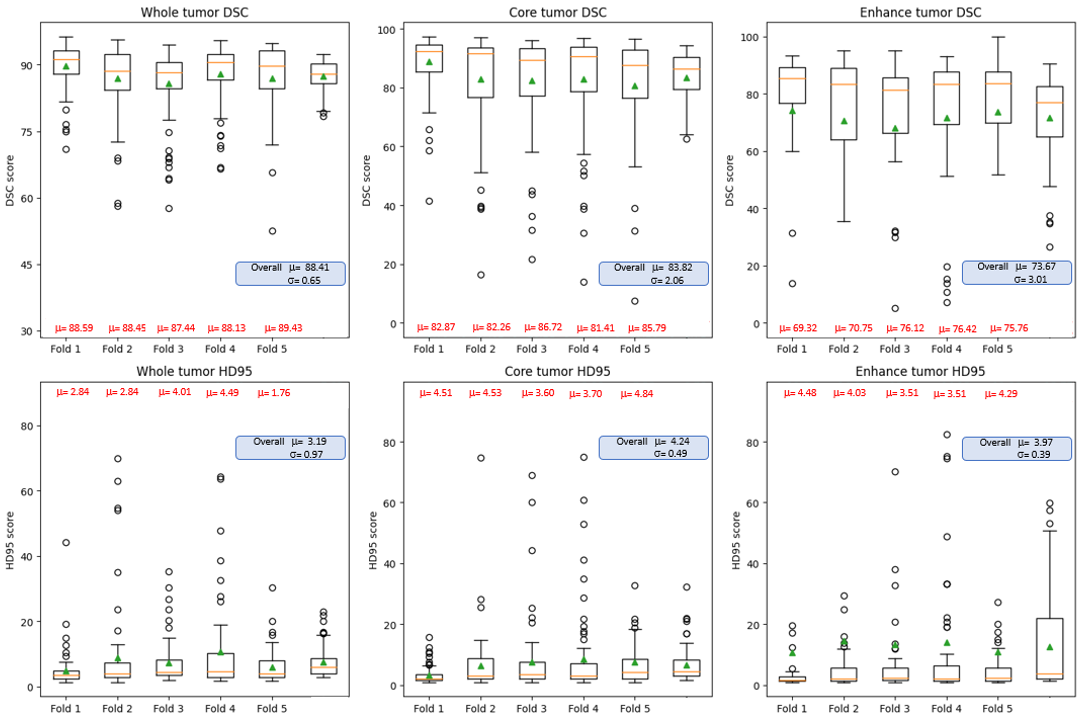}
  \caption{Boxplots of the DSC and HD95 metrics measured per sample (patient) on the BraTS~2020 five-fold cross-validation results with mean $\mu$ and standard deviation $\sigma$ using the LATUP-Net model. The orange line within each boxplot represents the median of the data. The green triangles represent the mean, and the circles denote the outliers. We also show the average distributions over all five folds. This figure provides a detailed view of our model's consistency and variability across samples.}\label{Figure:2020fold}
\end{figure*}

\begin{figure*}
  \centering
  \includegraphics[width=.9\textwidth]{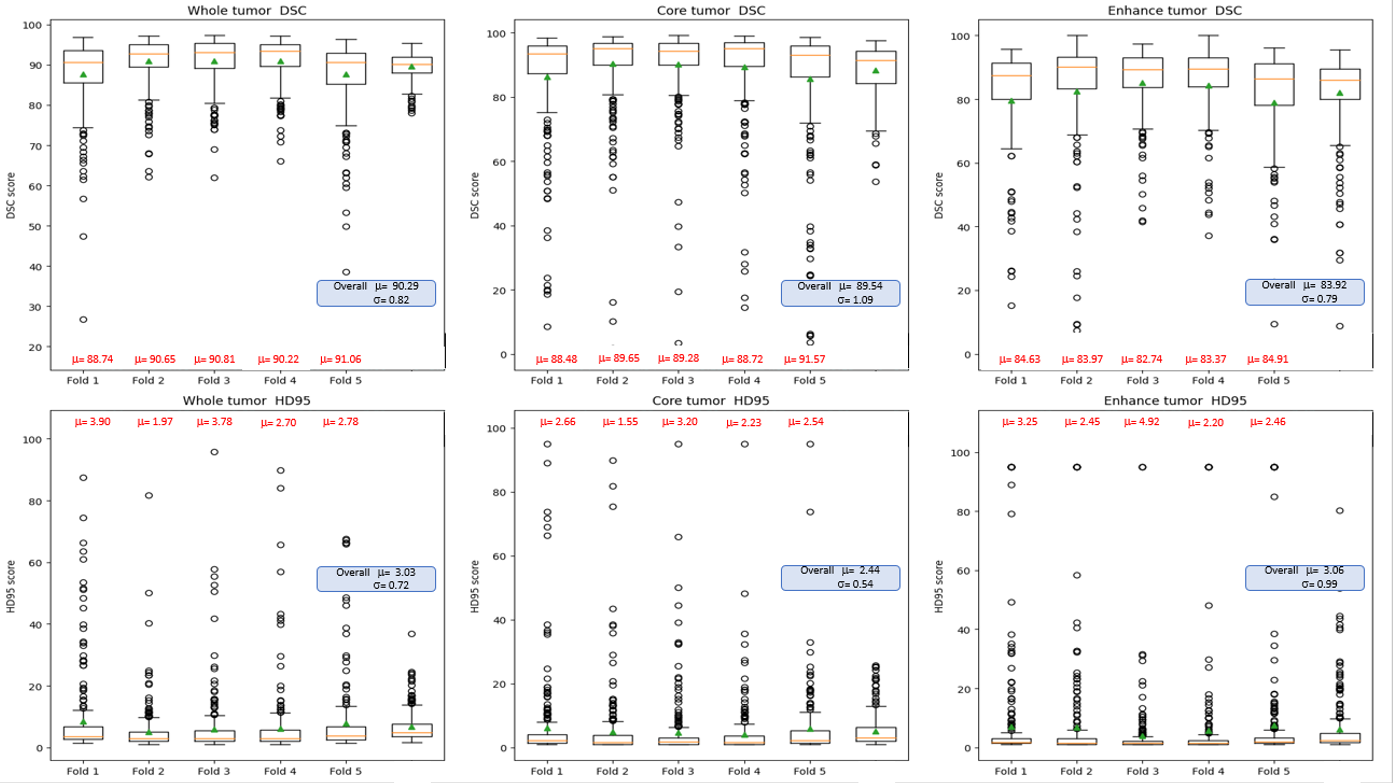}
  \caption{Boxplots of the DSC and HD95 metrics measured per sample (patient) on the BraTS~2021 five-fold cross-validation results with mean $\mu$ and standard deviation $\sigma$ using the LATUP-Net model. The orange line within each boxplot represents the median of the data. The green triangles represent the mean, and the circles denote the outliers. We also show the average distributions over all five folds. This figure provides a detailed view of our model's consistency and variability across samples.}\label{Figure:2021fold}
\end{figure*}

We perform five-fold cross-validation, with results shown in Figures~\ref{Figure:2020fold} and~\ref{Figure:2021fold}. These figures display the per-sample performance of our LATUP-Net model, highlighting the distribution of DSC and HD95 metrics, as well as variability, including outliers. Unlike other studies that report only average metrics, we provide a detailed analysis to demonstrate the model’s consistency across test samples.

We then compare the average results of our five-fold cross-validation with other high-performing models, as shown in Tables~\ref{table4_modified} and~\ref{table5_modified}. For these comparisons, we rely on the evaluation results reported in related publications, a common practice in brain tumor segmentation research. This approach is necessary because the source code for many existing methods is unavailable, and it helps avoid the bias that could be introduced by retraining models.

\subsubsection{Comparison with BraTS~2020 Results}

We compare our LATUP-Net model with various state-of-the-art models on the BraTS~2020 dataset. The evaluation focuses on three key metrics: the Dice similarity coefficient (DSC), the $95^\mathit{th}$ percentile Hausdorff distance (HD95), and the number of model parameters, alongside GFLOPs to reflect computational complexity. For our LATUP-Net model, we present the average performance across five-fold cross-validation to ensure a robust and comprehensive assessment. In contrast, for the state-of-the-art models, we report the results as they are presented in their respective original publications, which may include their best-split results or cross-validation outcomes. Detailed results of this comparative analysis are compiled in Table~\ref{table4_modified}.

Our LATUP-Net model demonstrates significant improvements in HD95 for all tumor regions (whole tumor, tumor core, and enhancing tumor), suggesting accurate predictions of tumor boundaries. This precision is of paramount importance in medical imaging since accurate boundaries can greatly impact clinical decision-making. However, it is worth noting that in the final evaluation of the five-fold cross-validation, the HD95 has been ignored for images that do not have clear boundaries, since having unclear, ambiguous ‘edges’ from which to measure the distances may mislead the results.

Our model surpasses several others in DSC measurements for whole tumor segmentation. Specifically, it outperforms Raza \emph{et al.}~\cite{raza2023dresu}, Ballestar \emph{et al.}~\cite{ballestar2020brain}, and Messaoudi \emph{et al.}~\cite{messaoudi2021efficient} by $1.81$, $4.2$, and $7.73$ respectively. Additionally, it achieves a notable improvement in efficiency, having only $3.07$~M parameters compared to $181.56$\cite{isensee2021nnu}, $32.99$\cite{wang2021transbts}, and $30.47$\cite{raza2023dresu}. Moreover, while Wang \emph{et al.}~\cite{wang2021transbts} model requires $333$ GFLOPs, and Raza \emph{et al.}~\cite{raza2023dresu} use $374.04$ GFLOPs, our model operates with only $15.79$ GFLOPs, highlighting its computational efficiency which is a crucial factor in real-world clinical applications.

When comparing the latest study by Zhu \emph{et al.}~\cite{zhu2024sparse}, which demonstrates high performance with a DSC of $90.22$ for the whole tumor, our model’s $88.41$ is still competitive. However, LATUP-Net surpasses Zhu's model in HD95, especially in tumor core and enhancing tumor predictions, demonstrating that it excels in boundary accuracy---a critical aspect in segmentation tasks. Although Zhu’s model outperforms in DSC across tumor regions, the computational complexity and parameter count are not mentioned. However, by incorporating transformer into the U-Net architecture, it is a moderately sized model compared to LATUP-Net, considering its performance trade-offs.

It is important to note that LATUP-Net's superior performance is not due to common confounding factors, such as ensembling or leveraging additional data. This aligns with the recent recommendations~\cite{isensee2024nnu}, which cautions against practices that artificially inflate a model's performance. Our architecture’s innovations---parallel convolution strategies and lightweight attention mechanisms---demonstrate genuine advancements in segmentation tasks without introducing unnecessary complexity or relying on inflated hardware resources. These results underscore the efficacy of simpler, well-configured models in achieving competitive performance in resource-constrained settings.

Although a deeper analysis is needed to fully determine the mechanisms behind this efficiency, the results from the earlier model comparison section provide concrete evidence of the effectiveness of our approach. Specifically, our adoption of parallel convolutions in the first encoder block seems to play a crucial role. This is evident as the PC model has $2.59$ million fewer parameters than U-Net, yet achieves rapid convergence during the initial training epochs, leading to shorter training times and reduced computational needs.

Brain tumor segmentation, particularly in the tumor core and enhancing tumor regions, poses significant challenges. While models like nnU-Net exhibit strong results, our LATUP-Net model achieves similar scores. It is critical to differentiate between DSC and HD95 scores. Although our HD95 scores indicate highly accurate boundary predictions, our DSC for the enhancing tumor (ET) is $73.67\%$, which is not the highest, highlighting an area for improvement in volumetric overlap with the ground truth.

In terms of the number of parameters, our model is remarkably efficient, with only $3.07$ million parameters, a stark contrast to nnU-Net’s $181.56$ million parameters. This efficiency reduces both computational costs and processing time, which is critical for real-time clinical applications, while maintaining comparable or even better performance in some aspects, as seen in the HD95 results. Although Raza \emph{et al.}~\cite{raza2023dresu} has a parameter count closer to ours, their performance does not match ours, underscoring the effectiveness of our architecture.

\begin{table*}[t]
\caption{Comparison of the performance and complexity of different methods on the BraTS~2020 test set. WT -- whole tumor, TC -- tumor core, ET -- enhancing tumor. Results highlighted in \textcolor{red}{red} indicate the best result, while those in \textcolor{blue}{blue} represent the second best. The symbol - indicates values not provided in the original paper.}\label{table4_modified}
\centering
\SetTblrInner{rowsep=3pt}
\resizebox{\textwidth}{!}{%
\begin{tblr}{
  colspec={Q[2.5cm,valign=m]llllllllll},
  cell{1-2}{1-9}={gray!20},
  cell{1}{1}={r=2}{l},
  cell{1}{2}={c=2}{l},
  cell{1}{4}={c=2}{l},
  cell{1}{6}={c=2}{l},
  cell{1}{8}={r=2}{l},
  cell{1}{9}={r=2}{l}
}
\hline
Study           & WT && TC && ET && Parameters & GFLOPs \\\cline{2-10}
                & DSC (\%) & HD95 (mm) & DSC (\%) & HD95 (mm) & DSC (\%) & HD95 (mm) & \\\hline
3D U-Net Baseline & $83.58$ & $18.50$ & $82.19$ & $15.38$ & $68.76$ & $19.34$ & $5.65$~M  & \textcolor{blue}{$23.08$} \\
Isensee \emph{et al.}~\cite{isensee2021nnu} & $88.95$ & $8.498$ & \textcolor{blue}{$85.06$} & $17.337$ & \textcolor{blue}{$82.03$} & $17.805$ & $181.56$~M & - \\
Tang \emph{et al.}~\cite{tang2021variational} & $89.29$ & $4.62$ & $78.97$ & $10.07$ & $70.30$ & $34.30$ & - & - \\
Ballestar \emph{et al.}~\cite{ballestar2020brain} & $84.21$ & $20.40$ & $75.03$ & $12.92$ & $61.75$ & $48.76$ & - & - \\
Wang \emph{et al.}~\cite{wang2021transbts} & \textcolor{blue}{$90.09$} & $4.96$ & $81.73$ & $9.76$ & $78.73$ & $17.94$ & $32.99$~M & 333 \\
Messaoudi \emph{et al.}~\cite{messaoudi2021efficient} & $80.68$ & - & $75.20$ & - & $69.59$ & - & - & - \\
Raza \emph{et al.}~\cite{raza2023dresu} & $86.60$ & - & $83.57$ & - & $80.04$ & - & $30.47$~M & 374.04 \\
Zhu \emph{et al.}~\cite{zhu2024sparse} & \textcolor{red}{$90.22$} & \textcolor{blue}{$4.03$} & \textcolor{red}{$89.20$} & \textcolor{red}{$3.30$} & \textcolor{red}{$82.48$} & \textcolor{red}{$2.29$} & - & - \\
LATUP-Net (proposed) & $88.41$ & \textcolor{red}{$3.19$} & $83.82$ & \textcolor{blue}{$4.24$} & $73.67$ & \textcolor{blue}{$3.97$} & \textcolor{red}{$3.07$}~M & \textcolor{red}{$15.79$} \\\hline
\end{tblr}}
\end{table*}

\begin{table*}[t]
\caption{Comparison of the performance and complexity of different methods on the BraTS~2021 test set. WT -- whole tumor, TC -- tumor core, ET -- enhancing tumor. Results highlighted in \textcolor{red}{red} indicate the best result, while those in \textcolor{blue}{blue} represent the second best. The symbol - indicates values not provided in the original paper.}\label{table5_modified}
\centering
\SetTblrInner{rowsep=3pt}
\resizebox{\textwidth}{!}{%
\begin{tblr}{
  colspec={Q[2.5cm,valign=m]llllllllll},
  cell{1-2}{1-9}={gray!20},
  cell{1}{1}={r=2}{l},
  cell{1}{2}={c=2}{l},
  cell{1}{4}={c=2}{l},
  cell{1}{6}={c=2}{l},
  cell{1}{8}={r=2}{l},
  cell{1}{9}={r=2}{l}
}
\hline
Study           & WT && TC && ET && Parameters & GFLOPs \\\cline{2-10}
                & DSC (\%) & HD95 (mm) & DSC (\%) & HD95 (mm) & DSC (\%) & HD95 (mm) & \\\hline
Peiris \emph{et al.}~\cite{peiris2021reciprocal} & $90.77$ & $5.37$ & $85.39$ & $8.5$ & $81.38$ & $21.83$ & - & - \\
Akbar \emph{et al.}~\cite{akbar2021unet3d} & $89.07$ & $11.78$ & $80.73$ & $21.17$ & $78.02$ & $25.8$ & - & - \\
Jia \emph{et al.}~\cite{jia2021hnf} & $92.53$ & \textcolor{blue}{$3.45$} & $87.96$ & $5.86$ & $84.80$ & $14.17$ & $17.91$~M & 449.79  \\
Li \emph{et al.}~\cite{li2021automatic} & $90.18$ & $6.15$ & $81.61$ & $16.65$ & $76.89$ & $30.21$ & - & - \\
Ma \emph{et al.}~\cite{ma2021nnunet} & $92.59$ & $3.80$ & $87.86$ & $9.20$ & $82.17$ & $21.09$ & - & - \\
Hatamizadeh \emph{et al.}~\cite{hatamizadeh2021swin} & \textcolor{blue}{$92.6$} & $5.83$ & $88.5$ & $3.77$ & \textcolor{blue}{$85.8$} & $6.01$ & $61.98$~M & \textcolor{blue}{$394.84$} \\
Roth \emph{et al.}~\cite{roth2021multi} & $90.6$ & $4.54$ & $83.5$ & $10.11$ & $79.2$ & $16.61$ & - & - \\
Zhu \emph{et al.}~\cite{zhu2024sparse} & \textcolor{red}{$93.10$} & $3.58$ & \textcolor{red}{$90.99$} & \textcolor{blue}{$3.27$} & \textcolor{red}{$87.64$} & \textcolor{red}{$2.57$} & - & - \\
LATUP-Net (proposed) & $90.29$ & \textcolor{red}{$3.03$} & \textcolor{blue}{$89.54$} & \textcolor{red}{$2.44$} & $83.92$ & \textcolor{blue}{$3.06$} & \textcolor{red}{$3.07$}~M & \textcolor{red}{$15.79$}\\\hline
\end{tblr}}
\end{table*}

\subsubsection{Comparison with BraTS~2021 Results}

Upon evaluating our LATUP-Net model on the BraTS 2021 dataset, it continues to show promising results, particularly in segmenting the tumor core, with a DSC of $89.54\%$, outperforming several models. Compared to Hatamizadeh \emph{et al.}~\cite{hatamizadeh2021swin}, who also achieved competitive results across tumor regions, LATUP-Net demonstrates a clear advantage in computational efficiency. Our model has only $3.07$ million parameters and requires $15.79$ GFLOPs, significantly less than Hatamizadeh's model, which requires $61.98$ million parameters and $394.84$ GFLOPs. This efficiency underscores the suitability of LATUP-Net for practical clinical use, especially in environments where computational resources may be limited.

In terms of HD95, LATUP-Net consistently achieves better results, with scores of $3.03$ for the whole tumor and $2.44$ for the tumor core.
Zhu \emph{et al.}~\cite{zhu2024sparse}, who achieved top DSC scores for the BraTS 2021 dataset, also produced competitive results in HD95. Nevertheless, LATUP-Net's reduced parameter count and GFLOPs stand out as key differentiators, especially for real-time clinical applications where faster processing times are essential.

Our model's DSC for the whole tumor and the enhancing tumor are $90.29\%$ and $83.92\%$, respectively. When we state that these scores are in line with leading models, we specifically refer to the work of Hatamizadeh \emph{et al.}, Jia \emph{et al.}~\cite{jia2021hnf}, and Ma \emph{et al.}~\cite{ma2021nnunet} 
(see Table~\ref{table5_modified}).

In summary, the LATUP-Net model offers a fine balance between segmentation accuracy and computational efficiency. Despite its compact architecture, it delivers strong performance, particularly in HD95 metrics, which are crucial for precise tumor boundary delineation. With only $3.07$ million parameters and $15.79$ GFLOPs, LATUP-Net is highly suited for resource-constrained environments, offering a feasible solution for real-time clinical deployment. While models like Zhu \emph{et al.}  surpass LATUP-Net in certain DSC scores, the efficiency and competitive performance across HD95 metrics highlight the real-world applicability of our model in brain tumor segmentation tasks.

\subsection{Limitations and Future Directions}\label{sec:limit}

Our model leverages multi-sequence MRI data for brain tumor segmentation and has demonstrated good performance on the BraTS~2020 and~2021 datasets, particularly when compared to other lightweight models working with similar data. However, we acknowledge that our work was trained on a relatively small dataset of about $1,200$ patients, which limits our understanding of the full covariance of potential data variations across different centers. This limitation is not unique to our study, as many segmentation models face similar challenges when dealing with multi-center data variations, especially regarding the availability of diverse data sources. We believe that access to larger, more heterogeneous datasets could help address these limitations, but this is an ongoing challenge within the field.

Other key limitations of our study lie in the consistency of model and parameter selection. While we have designed an effective architecture, the process of consistently selecting optimal models and fine-tuning hyperparameters has not been as rigorous as it could be. This challenge is partly due to the availability of computational resources, which restricts us from systematically testing a wider range of models and configurations. A more thorough exploration of model and parameter selection would be ideal, but this would require significant additional resources. As a result, we acknowledge that the model selection process could be improved and we plan to address this in future work.

Future work also includes adapting the architecture to other medical imaging segmentation tasks and refining the balance between attention and convolutional features, particularly to enhance our model's sensitivity to tumor regions and reduce variations in segmentation performance across different regions. Another crucial direction will be to explore the model's robustness when faced with incomplete or missing MRI modalities, a common scenario in real-world clinical settings. Ensuring that LATUP-Net can maintain high segmentation performance even when some modalities are unavailable will enhance its practical applicability in diverse medical environments. Additionally, we are currently studying the robustness and explainability of our model, which are critical for clinical applications~\cite{alwadee2024}.
However, this ongoing research extends beyond the scope of the present paper and will be addressed in future studies.

\section{Conclusion}\label{sec:conclusion}

In this work, we have unveiled the LATUP-Net network, an enhanced U-Net variant for 3D brain tumor segmentation designed to be lightweight in its computational demand. This model substantially decreases the number of parameters needed while maintaining, and in some aspects surpassing, the segmentation performance of state-of-the-art methods. With $3.07$~M parameters, about $59$ times fewer parameters than the state-of-the-art nnU-Net with $181.56$~M parameters, LATUP-Net underscores an advancement where efficient modeling coupled with parallel convolutions can lead to a significant reduction in overfitting risk and more judicious use of computational resources.

Our model demonstrates an impressive ability to delineate tumor boundaries with high accuracy, as evidenced by its performance in Hausdorff distance (HD95) measurements. These achievements indicate the model's potential utility in clinical settings, where precise segmentations are integral to formulating effective treatment plans. Furthermore, LATUP-Net's lightweight architecture, requiring only $15.79$~GFLOPs, makes it particularly suitable for deployment in resource-constrained environments, such as developing countries, where computational resources may be limited.

A pivotal aspect of our research is incorporating attention mechanisms, which refine our model's capability to focus on salient features within MRI scans. Our comparative analysis across different attention mechanisms, such as SE, CBAM, and ECA, reveals that while all contribute to accuracy improvements, SE provides a balance between performance and parameter efficiency, particularly in delineating enhanced tumors. However, the enhancements brought by attention are found to be nuanced. The slight underperformance in Dice score coefficients for enhancing tumor segmentation suggests that attention mechanisms do not unilaterally enhance performance across all regions. This is corroborated by gradient-weighted class activation mapping (Grad-CAM) and confusion matrix analyses. These investigations highlight scenarios where attention mechanisms seem to focus too narrowly on local features, occasionally at the expense of contextual understanding, leading to potential misclassification between regions with texturally similar features. The attention-enhanced model, while showing promise in segmenting small regions, also illustrates that there are instances where traditional convolutions may suffice and that the features they capture can be integral to achieving precise segmentations.

The LATUP-Net model stands as a testament to the possibility of achieving state-of-the-art performance with a fraction of the computational cost, highlighting a promising direction for medical image analysis research and the development of practical, accessible tools for brain tumor segmentation. In real-world clinical applications, however, the dependency on multi-sequence MRI data presents a practical challenge, as some modalities may be unavailable in certain settings. Addressing this limitation is a key consideration for future work.

\printcredits

\bibliographystyle{model1-num-names}
\bibliography{ccas-refs.bib}

\begin{thebibliography}{64}
\expandafter\ifx\csname natexlab\endcsname\relax\def\natexlab#1{#1}\fi
\providecommand{\url}[1]{\texttt{#1}}
\providecommand{\href}[2]{#2}
\providecommand{\path}[1]{#1}
\providecommand{\DOIprefix}{doi:}
\providecommand{\ArXivprefix}{arXiv:}
\providecommand{\URLprefix}{URL: }
\providecommand{\Pubmedprefix}{pmid:}
\providecommand{\doi}[1]{\href{http://dx.doi.org/#1}{\path{#1}}}
\providecommand{\Pubmed}[1]{\href{pmid:#1}{\path{#1}}}
\providecommand{\bibinfo}[2]{#2}
\ifx\xfnm\relax \def\xfnm[#1]{\unskip,\space#1}\fi
\bibitem[{Menze et~al.(2014)Menze, Jakab, Bauer, Kalpathy-Cramer, Farahani,
  Kirby, Burren, Porz, Slotboom, Wiest et~al.}]{menze2014multimodal}
\bibinfo{author}{B.~H. Menze}, \bibinfo{author}{A.~Jakab},
  \bibinfo{author}{S.~Bauer}, \bibinfo{author}{J.~Kalpathy-Cramer},
  \bibinfo{author}{K.~Farahani}, \bibinfo{author}{J.~Kirby},
  \bibinfo{author}{Y.~Burren}, \bibinfo{author}{N.~Porz},
  \bibinfo{author}{J.~Slotboom}, \bibinfo{author}{R.~Wiest}, et~al.,
\newblock \bibinfo{title}{The multimodal brain tumor image segmentation
  benchmark {(BRATS)}},
\newblock \bibinfo{journal}{IEEE Transactions on Medical Imaging}
  \bibinfo{volume}{34} (\bibinfo{year}{2014}) \bibinfo{pages}{1993--2024}.
\bibitem[{I{\c{s}}{\i}n et~al.(2016)I{\c{s}}{\i}n, Direko{\u{g}}lu, and
  {\c{S}}ah}]{icsin2016review}
\bibinfo{author}{A.~I{\c{s}}{\i}n}, \bibinfo{author}{C.~Direko{\u{g}}lu},
  \bibinfo{author}{M.~{\c{S}}ah},
\newblock \bibinfo{title}{Review of {MRI-based} brain tumor image segmentation
  using deep learning methods},
\newblock \bibinfo{journal}{Procedia Computer Science} \bibinfo{volume}{102}
  (\bibinfo{year}{2016}) \bibinfo{pages}{317--324}.
\bibitem[{Bakas et~al.(2017)Bakas, Akbari, Sotiras, Bilello, Rozycki, Kirby,
  Freymann, Farahani, and Davatzikos}]{bakas2017advancing}
\bibinfo{author}{S.~Bakas}, \bibinfo{author}{H.~Akbari},
  \bibinfo{author}{A.~Sotiras}, \bibinfo{author}{M.~Bilello},
  \bibinfo{author}{M.~Rozycki}, \bibinfo{author}{J.~S. Kirby},
  \bibinfo{author}{J.~B. Freymann}, \bibinfo{author}{K.~Farahani},
  \bibinfo{author}{C.~Davatzikos},
\newblock \bibinfo{title}{Advancing the cancer genome atlas glioma {MRI}
  collections with expert segmentation labels and radiomic features},
\newblock \bibinfo{journal}{Scientific data} \bibinfo{volume}{4}
  (\bibinfo{year}{2017}) \bibinfo{pages}{1--13}.
\bibitem[{Bakas et~al.(2018)Bakas, Reyes, Jakab, Bauer, Rempfler, Crimi,
  Shinohara, Berger, Ha, Rozycki et~al.}]{bakas2018identifying}
\bibinfo{author}{S.~Bakas}, \bibinfo{author}{M.~Reyes},
  \bibinfo{author}{A.~Jakab}, \bibinfo{author}{S.~Bauer},
  \bibinfo{author}{M.~Rempfler}, \bibinfo{author}{A.~Crimi},
  \bibinfo{author}{R.~T. Shinohara}, \bibinfo{author}{C.~Berger},
  \bibinfo{author}{S.~M. Ha}, \bibinfo{author}{M.~Rozycki}, et~al.,
\newblock \bibinfo{title}{Identifying the best machine learning algorithms for
  brain tumor segmentation, progression assessment, and overall survival
  prediction in the brats challenge},
\newblock \bibinfo{journal}{arXiv:1811.02629}  (\bibinfo{year}{2018}).
\bibitem[{Li et~al.(2019)Li, Li, and Wang}]{li2019novel}
\bibinfo{author}{H.~Li}, \bibinfo{author}{A.~Li}, \bibinfo{author}{M.~Wang},
\newblock \bibinfo{title}{A novel end-to-end brain tumor segmentation method
  using improved fully convolutional networks},
\newblock \bibinfo{journal}{Computers in Biology and Medicine}
  \bibinfo{volume}{108} (\bibinfo{year}{2019}) \bibinfo{pages}{150--160}.
\bibitem[{Wieczorek et~al.(2021)Wieczorek, Si{\l}ka, Wo{\'z}niak, Garg, and
  Hassan}]{wieczorek2021lightweight}
\bibinfo{author}{M.~Wieczorek}, \bibinfo{author}{J.~Si{\l}ka},
  \bibinfo{author}{M.~Wo{\'z}niak}, \bibinfo{author}{S.~Garg},
  \bibinfo{author}{M.~M. Hassan},
\newblock \bibinfo{title}{Lightweight convolutional neural network model for
  human face detection in risk situations},
\newblock \bibinfo{journal}{IEEE Transactions on Industrial Informatics}
  \bibinfo{volume}{18} (\bibinfo{year}{2021}) \bibinfo{pages}{4820--4829}.
\bibitem[{Wo{\'z}niak et~al.(2021)Wo{\'z}niak, Si{\l}ka, and
  Wieczorek}]{wozniak2021deep}
\bibinfo{author}{M.~Wo{\'z}niak}, \bibinfo{author}{J.~Si{\l}ka},
  \bibinfo{author}{M.~Wieczorek},
\newblock \bibinfo{title}{Deep neural network correlation learning mechanism
  for ct brain tumor detection},
\newblock \bibinfo{journal}{Neural Computing and Applications}
  (\bibinfo{year}{2021}) \bibinfo{pages}{1--16}.
\bibitem[{Ronneberger et~al.(2015)Ronneberger, Fischer, and
  Brox}]{ronneberger2015u}
\bibinfo{author}{O.~Ronneberger}, \bibinfo{author}{P.~Fischer},
  \bibinfo{author}{T.~Brox},
\newblock \bibinfo{title}{{U-Net}: Convolutional networks for biomedical image
  segmentation},
\newblock in: \bibinfo{booktitle}{Medical Image Computing and Computer-Assisted
  Intervention--MICCAI: 18th International Conference, Proceedings, Part III
  18}, \bibinfo{organization}{Springer}, \bibinfo{year}{2015}, pp.
  \bibinfo{pages}{234--241}.
\bibitem[{Isensee et~al.(2024)Isensee, Wald, Ulrich, Baumgartner, Roy,
  Maier-Hein, and Jaeger}]{isensee2024nnu}
\bibinfo{author}{F.~Isensee}, \bibinfo{author}{T.~Wald},
  \bibinfo{author}{C.~Ulrich}, \bibinfo{author}{M.~Baumgartner},
  \bibinfo{author}{S.~Roy}, \bibinfo{author}{K.~Maier-Hein},
  \bibinfo{author}{P.~F. Jaeger},
\newblock \bibinfo{title}{{nnU-Net} revisited: A call for rigorous validation
  in 3d medical image segmentation},
\newblock \bibinfo{journal}{arXiv preprint arXiv:2404.09556}
  (\bibinfo{year}{2024}).
\bibitem[{Zhu et~al.(2024)Zhu, Sun, Qi, Li, Gao, and Liu}]{zhu2024sparse}
\bibinfo{author}{Z.~Zhu}, \bibinfo{author}{M.~Sun}, \bibinfo{author}{G.~Qi},
  \bibinfo{author}{Y.~Li}, \bibinfo{author}{X.~Gao}, \bibinfo{author}{Y.~Liu},
\newblock \bibinfo{title}{{Sparse Dynamic Volume TransUNet} with multi-level
  edge fusion for brain tumor segmentation},
\newblock \bibinfo{journal}{Computers in Biology and Medicine}
  (\bibinfo{year}{2024}) \bibinfo{pages}{108284}.
\bibitem[{Xu et~al.(2024)Xu, Yu, Qi, Gong, Qu, Yin, and Yang}]{xu2024brain}
\bibinfo{author}{Y.~Xu}, \bibinfo{author}{K.~Yu}, \bibinfo{author}{G.~Qi},
  \bibinfo{author}{Y.~Gong}, \bibinfo{author}{X.~Qu}, \bibinfo{author}{L.~Yin},
  \bibinfo{author}{P.~Yang},
\newblock \bibinfo{title}{Brain tumour segmentation framework with deep nuanced
  reasoning and swin-t},
\newblock \bibinfo{journal}{IET Image Processing} \bibinfo{volume}{18}
  (\bibinfo{year}{2024}) \bibinfo{pages}{1550--1564}.
\bibitem[{Wu et~al.(2022)Wu, Hu, and Liu}]{wu2022mr}
\bibinfo{author}{L.~Wu}, \bibinfo{author}{S.~Hu}, \bibinfo{author}{C.~Liu},
\newblock \bibinfo{title}{{MR} brain segmentation based on {DE-ResUnet}
  combining texture features and background knowledge},
\newblock \bibinfo{journal}{Biomedical Signal Processing and Control}
  \bibinfo{volume}{75} (\bibinfo{year}{2022}) \bibinfo{pages}{103541}.
\bibitem[{D~R~Sarvamangala(2022)}]{sarvamangala2022convolutional}
\bibinfo{author}{R.~V.~K. D~R~Sarvamangala},
\newblock \bibinfo{title}{Convolutional neural networks in medical image
  understanding: a survey},
\newblock \bibinfo{journal}{Evolutionary Intelligence} \bibinfo{volume}{15}
  (\bibinfo{year}{2022}) \bibinfo{pages}{1--22}.
\bibitem[{Beutel(2000)}]{beutel2000handbook}
\bibinfo{author}{J.~Beutel}, \bibinfo{title}{Handbook of Medical Imaging},
  volume~\bibinfo{volume}{3}, \bibinfo{publisher}{Spie Press},
  \bibinfo{year}{2000}.
\bibitem[{Ma et~al.(2023)Ma, Wang, and Hu}]{ma2023lmu}
\bibinfo{author}{T.~Ma}, \bibinfo{author}{K.~Wang}, \bibinfo{author}{F.~Hu},
\newblock \bibinfo{title}{{LMU-Net:} lightweight u-shaped network for medical
  image segmentation},
\newblock \bibinfo{journal}{Medical \& biological engineering \& computing}
  (\bibinfo{year}{2023}) \bibinfo{pages}{1--10}.
\bibitem[{Vaswani(2017)}]{vaswani2017attention}
\bibinfo{author}{A.~Vaswani},
\newblock \bibinfo{title}{Attention is all you need},
\newblock \bibinfo{journal}{Advances in Neural Information Processing Systems}
  (\bibinfo{year}{2017}).
\bibitem[{Park(2018)}]{park2018bam}
\bibinfo{author}{J.~Park},
\newblock \bibinfo{title}{Bam: Bottleneck attention module},
\newblock \bibinfo{journal}{arXiv preprint arXiv:1807.06514}
  (\bibinfo{year}{2018}).
\bibitem[{Woo et~al.(2018)Woo, Park, Lee, and Kweon}]{woo2018cbam}
\bibinfo{author}{S.~Woo}, \bibinfo{author}{J.~Park}, \bibinfo{author}{J.-Y.
  Lee}, \bibinfo{author}{I.~S. Kweon},
\newblock \bibinfo{title}{{CBAM}: Convolutional block attention module},
\newblock in: \bibinfo{booktitle}{Proceedings of the European Conference on
  Computer Vision (ECCV)}, \bibinfo{year}{2018}, pp. \bibinfo{pages}{3--19}.
\bibitem[{Hu et~al.(2018)Hu, Shen, and Sun}]{hu2018squeeze}
\bibinfo{author}{J.~Hu}, \bibinfo{author}{L.~Shen}, \bibinfo{author}{G.~Sun},
\newblock \bibinfo{title}{Squeeze-and-excitation networks},
\newblock in: \bibinfo{booktitle}{Proceedings of the IEEE Conference on
  Computer Vision and Pattern Recognition}, \bibinfo{year}{2018}, pp.
  \bibinfo{pages}{7132--7141}.
\bibitem[{Selvaraju et~al.(2017)Selvaraju, Cogswell, Das, Vedantam, Parikh, and
  Batra}]{selvaraju2017grad}
\bibinfo{author}{R.~R. Selvaraju}, \bibinfo{author}{M.~Cogswell},
  \bibinfo{author}{A.~Das}, \bibinfo{author}{R.~Vedantam},
  \bibinfo{author}{D.~Parikh}, \bibinfo{author}{D.~Batra},
\newblock \bibinfo{title}{Grad-cam: Visual explanations from deep networks via
  gradient-based localization},
\newblock in: \bibinfo{booktitle}{Proceedings of the IEEE International
  Conference on Computer Vision}, \bibinfo{year}{2017}, pp.
  \bibinfo{pages}{618--626}.
\bibitem[{{\c{C}}i{\c{c}}ek et~al.(2016){\c{C}}i{\c{c}}ek, Abdulkadir,
  Lienkamp, Brox, and Ronneberger}]{cciccek20163d}
\bibinfo{author}{{\"O}.~{\c{C}}i{\c{c}}ek}, \bibinfo{author}{A.~Abdulkadir},
  \bibinfo{author}{S.~S. Lienkamp}, \bibinfo{author}{T.~Brox},
  \bibinfo{author}{O.~Ronneberger},
\newblock \bibinfo{title}{{3D U-Net:} learning dense volumetric segmentation
  from sparse annotation},
\newblock in: \bibinfo{booktitle}{Medical Image Computing and Computer-Assisted
  Intervention--MICCAI: 19th International Conference, Proceedings, Part II
  19}, \bibinfo{organization}{Springer}, \bibinfo{year}{2016}, pp.
  \bibinfo{pages}{424--432}.
\bibitem[{Chen et~al.(2019)Chen, Liu, Peng, Sun, and Qiao}]{chen2019s3d}
\bibinfo{author}{W.~Chen}, \bibinfo{author}{B.~Liu}, \bibinfo{author}{S.~Peng},
  \bibinfo{author}{J.~Sun}, \bibinfo{author}{X.~Qiao},
\newblock \bibinfo{title}{{S3D-UNet}: separable {3D U-Net} for brain tumor
  segmentation},
\newblock in: \bibinfo{booktitle}{Brainlesion: Glioma, Multiple Sclerosis,
  Stroke and Traumatic Brain Injuries: 4th International Workshop, Part II 4},
  \bibinfo{organization}{Springer}, \bibinfo{year}{2019}, pp.
  \bibinfo{pages}{358--368}.
\bibitem[{Wang et~al.(2021)Wang, Chen, Ding, Yu, Zha, and
  Li}]{wang2021transbts}
\bibinfo{author}{W.~Wang}, \bibinfo{author}{C.~Chen},
  \bibinfo{author}{M.~Ding}, \bibinfo{author}{H.~Yu}, \bibinfo{author}{S.~Zha},
  \bibinfo{author}{J.~Li},
\newblock \bibinfo{title}{{TransBTS}: Multimodal brain tumor segmentation using
  transformer},
\newblock in: \bibinfo{booktitle}{Medical Image Computing and Computer Assisted
  Intervention--MICCAI: 24th International Conference, Proceedings, Part I 24},
  \bibinfo{organization}{Springer}, \bibinfo{year}{2021}, pp.
  \bibinfo{pages}{109--119}.
\bibitem[{Zhu et~al.(2023)Zhu, He, Qi, Li, Cong, and Liu}]{zhu2023brain}
\bibinfo{author}{Z.~Zhu}, \bibinfo{author}{X.~He}, \bibinfo{author}{G.~Qi},
  \bibinfo{author}{Y.~Li}, \bibinfo{author}{B.~Cong}, \bibinfo{author}{Y.~Liu},
\newblock \bibinfo{title}{Brain tumor segmentation based on the fusion of deep
  semantics and edge information in multimodal {MRI}},
\newblock \bibinfo{journal}{Information Fusion} \bibinfo{volume}{91}
  (\bibinfo{year}{2023}) \bibinfo{pages}{376--387}.
\bibitem[{Liu et~al.(2019)Liu, Zhang, Cai, Chen, Yun, Feng, and
  Yang}]{liu2019automatic}
\bibinfo{author}{Y.~Liu}, \bibinfo{author}{X.~Zhang}, \bibinfo{author}{G.~Cai},
  \bibinfo{author}{Y.~Chen}, \bibinfo{author}{Z.~Yun},
  \bibinfo{author}{Q.~Feng}, \bibinfo{author}{W.~Yang},
\newblock \bibinfo{title}{Automatic delineation of ribs and clavicles in chest
  radiographs using fully convolutional {DenseNets}},
\newblock \bibinfo{journal}{Computer Methods and Programs in Biomedicine}
  \bibinfo{volume}{180} (\bibinfo{year}{2019}) \bibinfo{pages}{105014}.
\bibitem[{Oh et~al.(2019)Oh, Ng, San~Tan, and Acharya}]{oh2019automated}
\bibinfo{author}{S.~L. Oh}, \bibinfo{author}{E.~Y. Ng},
  \bibinfo{author}{R.~San~Tan}, \bibinfo{author}{U.~R. Acharya},
\newblock \bibinfo{title}{Automated beat-wise arrhythmia diagnosis using
  modified {U-Net} on extended electrocardiographic recordings with
  heterogeneous arrhythmia types},
\newblock \bibinfo{journal}{Computers in Biology and Medicine}
  \bibinfo{volume}{105} (\bibinfo{year}{2019}) \bibinfo{pages}{92--101}.
\bibitem[{Liu et~al.(2019)Liu, Song, Sheng, Wang, Jiang, Zhang, and
  Yuan}]{liu2019liver}
\bibinfo{author}{Z.~Liu}, \bibinfo{author}{Y.-Q. Song}, \bibinfo{author}{V.~S.
  Sheng}, \bibinfo{author}{L.~Wang}, \bibinfo{author}{R.~Jiang},
  \bibinfo{author}{X.~Zhang}, \bibinfo{author}{D.~Yuan},
\newblock \bibinfo{title}{Liver {CT} sequence segmentation based with improved
  {U-Net} and graph cut},
\newblock \bibinfo{journal}{Expert Systems with Applications}
  \bibinfo{volume}{126} (\bibinfo{year}{2019}) \bibinfo{pages}{54--63}.
\bibitem[{Zhang et~al.(2020)Zhang, Wu, Coleman, and Kerr}]{zhang2020dense}
\bibinfo{author}{Z.~Zhang}, \bibinfo{author}{C.~Wu},
  \bibinfo{author}{S.~Coleman}, \bibinfo{author}{D.~Kerr},
\newblock \bibinfo{title}{{DENSE-INception U-Net} for medical image
  segmentation},
\newblock \bibinfo{journal}{Computer Methods and Programs in Biomedicine}
  \bibinfo{volume}{192} (\bibinfo{year}{2020}) \bibinfo{pages}{105395}.
\bibitem[{Chen et~al.(2019)Chen, Liu, Ding, Zheng, and Li}]{chen20193d}
\bibinfo{author}{C.~Chen}, \bibinfo{author}{X.~Liu}, \bibinfo{author}{M.~Ding},
  \bibinfo{author}{J.~Zheng}, \bibinfo{author}{J.~Li},
\newblock \bibinfo{title}{{3D} dilated multi-fiber network for real-time brain
  tumor segmentation in {MRI}},
\newblock in: \bibinfo{booktitle}{Medical Image Computing and Computer Assisted
  Intervention--MICCAI: 22nd International Conference, Proceedings, Part III
  22}, \bibinfo{organization}{Springer}, \bibinfo{year}{2019}, pp.
  \bibinfo{pages}{184--192}.
\bibitem[{Luo et~al.(2020)Luo, Jia, Yuan, and Peng}]{luo2020hdc}
\bibinfo{author}{Z.~Luo}, \bibinfo{author}{Z.~Jia}, \bibinfo{author}{Z.~Yuan},
  \bibinfo{author}{J.~Peng},
\newblock \bibinfo{title}{{HDC-Net:} hierarchical decoupled convolution network
  for brain tumor segmentation},
\newblock \bibinfo{journal}{IEEE Journal of Biomedical and Health Informatics}
  \bibinfo{volume}{25} (\bibinfo{year}{2020}) \bibinfo{pages}{737--745}.
\bibitem[{Magadza and Viriri(2022)}]{magadza2022brain}
\bibinfo{author}{T.~Magadza}, \bibinfo{author}{S.~Viriri},
\newblock \bibinfo{title}{Brain tumor segmentation using partial depthwise
  separable convolutions},
\newblock \bibinfo{journal}{IEEE Access} \bibinfo{volume}{10}
  (\bibinfo{year}{2022}) \bibinfo{pages}{124206--124216}.
\bibitem[{Zhu et~al.(2024)Zhu, Wang, Qi, Mazur, Yang, and Liu}]{zhu2024brain}
\bibinfo{author}{Z.~Zhu}, \bibinfo{author}{Z.~Wang}, \bibinfo{author}{G.~Qi},
  \bibinfo{author}{N.~Mazur}, \bibinfo{author}{P.~Yang},
  \bibinfo{author}{Y.~Liu},
\newblock \bibinfo{title}{Brain tumor segmentation in {MRI} with multi-modality
  spatial information enhancement and boundary shape correction},
\newblock \bibinfo{journal}{Pattern Recognition} \bibinfo{volume}{153}
  (\bibinfo{year}{2024}) \bibinfo{pages}{110553}.
\bibitem[{Roy et~al.(2018)Roy, Navab, and Wachinger}]{roy2018concurrent}
\bibinfo{author}{A.~G. Roy}, \bibinfo{author}{N.~Navab},
  \bibinfo{author}{C.~Wachinger},
\newblock \bibinfo{title}{Concurrent spatial and channel ‘squeeze \&
  excitation’in fully convolutional networks},
\newblock in: \bibinfo{booktitle}{Medical Image Computing and Computer Assisted
  Intervention--MICCAI: 21st International Conference, Proceedings, Part I},
  \bibinfo{organization}{Springer}, \bibinfo{year}{2018}, pp.
  \bibinfo{pages}{421--429}.
\bibitem[{Ulyanov et~al.(2016)Ulyanov, Vedaldi, and
  Lempitsky}]{ulyanov2016instance}
\bibinfo{author}{D.~Ulyanov}, \bibinfo{author}{A.~Vedaldi},
  \bibinfo{author}{V.~Lempitsky},
\newblock \bibinfo{title}{Instance normalization: The missing ingredient for
  fast stylization},
\newblock \bibinfo{journal}{arXiv:1607.08022}  (\bibinfo{year}{2016}).
\bibitem[{Alwadee and Langbein(2024)}]{bca}
\bibinfo{author}{E.~Alwadee}, \bibinfo{author}{F.~C. Langbein},
  \bibinfo{title}{Bca - brain cancer segmentation python package, version 1.0,
  software}, \bibinfo{year}{2024}. \URLprefix
  \url{https://qyber.black/ca/code-bca}.
\bibitem[{Rajamani et~al.(2023)Rajamani, Rani, Siebert, ElagiriRamalingam, and
  Heinrich}]{rajamani2023attention}
\bibinfo{author}{K.~T. Rajamani}, \bibinfo{author}{P.~Rani},
  \bibinfo{author}{H.~Siebert}, \bibinfo{author}{R.~ElagiriRamalingam},
  \bibinfo{author}{M.~P. Heinrich},
\newblock \bibinfo{title}{Attention-augmented {U-Net (AA-U-Net)} for semantic
  segmentation},
\newblock \bibinfo{journal}{Signal, image and video processing}
  \bibinfo{volume}{17} (\bibinfo{year}{2023}) \bibinfo{pages}{981--989}.
\bibitem[{Chen(2014)}]{chen2014research}
\bibinfo{author}{X.~Chen},
\newblock \bibinfo{title}{Research on algorithm and application of deep
  learning based on convolutional neural network},
\newblock \bibinfo{journal}{Zhejiang Gongshang University}
  (\bibinfo{year}{2014}).
\bibitem[{Szegedy et~al.(2015)Szegedy, Liu, Jia, Sermanet, Reed, Anguelov,
  Erhan, Vanhoucke, and Rabinovich}]{szegedy2015going}
\bibinfo{author}{C.~Szegedy}, \bibinfo{author}{W.~Liu},
  \bibinfo{author}{Y.~Jia}, \bibinfo{author}{P.~Sermanet},
  \bibinfo{author}{S.~Reed}, \bibinfo{author}{D.~Anguelov},
  \bibinfo{author}{D.~Erhan}, \bibinfo{author}{V.~Vanhoucke},
  \bibinfo{author}{A.~Rabinovich},
\newblock \bibinfo{title}{Going deeper with convolutions},
\newblock in: \bibinfo{booktitle}{Proceedings of the IEEE Conference on
  Computer Vision and Pattern Recognition}, \bibinfo{year}{2015}, pp.
  \bibinfo{pages}{1--9}.
\bibitem[{Wang et~al.(2020)Wang, Wu, Zhu, Li, Zuo, and Hu}]{wang2020eca}
\bibinfo{author}{Q.~Wang}, \bibinfo{author}{B.~Wu}, \bibinfo{author}{P.~Zhu},
  \bibinfo{author}{P.~Li}, \bibinfo{author}{W.~Zuo}, \bibinfo{author}{Q.~Hu},
\newblock \bibinfo{title}{{ECA-Net}: Efficient channel attention for deep
  convolutional neural networks},
\newblock in: \bibinfo{booktitle}{Proceedings of the IEEE/CVF Conference on
  Computer Vision and Pattern Recognition}, \bibinfo{year}{2020}, pp.
  \bibinfo{pages}{11534--11542}.
\bibitem[{Gu et~al.(2019)Gu, Sun, Zhang, Fu, and Wang}]{gu2019deep}
\bibinfo{author}{J.~Gu}, \bibinfo{author}{X.~Sun}, \bibinfo{author}{Y.~Zhang},
  \bibinfo{author}{K.~Fu}, \bibinfo{author}{L.~Wang},
\newblock \bibinfo{title}{Deep residual squeeze and excitation network for
  remote sensing image super-resolution},
\newblock \bibinfo{journal}{Remote Sensing} \bibinfo{volume}{11}
  (\bibinfo{year}{2019}) \bibinfo{pages}{1817}.
\bibitem[{Raschka(2018)}]{raschka2018model}
\bibinfo{author}{S.~Raschka},
\newblock \bibinfo{title}{Model evaluation, model selection, and algorithm
  selection in machine learning},
\newblock \bibinfo{journal}{arXiv:1811.12808}  (\bibinfo{year}{2018}).
\bibitem[{Patro and Sahu(2015)}]{patro2015normalization}
\bibinfo{author}{S.~Patro}, \bibinfo{author}{K.~K. Sahu},
\newblock \bibinfo{title}{Normalization: A preprocessing stage},
\newblock \bibinfo{journal}{arXiv:1503.06462}  (\bibinfo{year}{2015}).
\bibitem[{Kingma and Ba(2014)}]{kingma2014adam}
\bibinfo{author}{D.~P. Kingma}, \bibinfo{author}{J.~Ba},
\newblock \bibinfo{title}{Adam: A method for stochastic optimization},
\newblock \bibinfo{journal}{arXiv:1412.6980}  (\bibinfo{year}{2014}).
\bibitem[{Isensee et~al.(2021)Isensee, J{\"a}ger, Full, Vollmuth, and
  Maier-Hein}]{isensee2021nnu}
\bibinfo{author}{F.~Isensee}, \bibinfo{author}{P.~F. J{\"a}ger},
  \bibinfo{author}{P.~M. Full}, \bibinfo{author}{P.~Vollmuth},
  \bibinfo{author}{K.~H. Maier-Hein},
\newblock \bibinfo{title}{{nnU-Net} for brain tumor segmentation},
\newblock in: \bibinfo{booktitle}{Brainlesion: Glioma, Multiple Sclerosis,
  Stroke and Traumatic Brain Injuries: 6th International Workshop, BrainLes
  2020, Part II 6}, \bibinfo{organization}{Springer}, \bibinfo{year}{2021}, pp.
  \bibinfo{pages}{118--132}.
\bibitem[{Maas et~al.(2013)Maas, Hannun, Ng et~al.}]{maas2013rectifier}
\bibinfo{author}{A.~L. Maas}, \bibinfo{author}{A.~Y. Hannun},
  \bibinfo{author}{A.~Y. Ng}, et~al.,
\newblock \bibinfo{title}{Rectifier nonlinearities improve neural network
  acoustic models},
\newblock in: \bibinfo{booktitle}{Proc. {ICML}}, volume~\bibinfo{volume}{28},
  \bibinfo{year}{2013}, p.~\bibinfo{pages}{3}.
\bibitem[{Alwadee and Langbein(2024)}]{bca-results}
\bibinfo{author}{E.~Alwadee}, \bibinfo{author}{F.~C. Langbein},
  \bibinfo{title}{{BCa} segmentation results: {LATUPNet}, version 1.0. software
  and data}, \bibinfo{year}{2024}. \URLprefix
  \url{https://qyber.black/ca/results-bca-latup}.
\bibitem[{Bakas et~al.(2017)Bakas, Akbari, Sotiras, Bilello, Rozycki, Kirby,
  Freymann, Farahani, and Davatzikos}]{bakas2017segmentation}
\bibinfo{author}{S.~Bakas}, \bibinfo{author}{H.~Akbari},
  \bibinfo{author}{A.~Sotiras}, \bibinfo{author}{M.~Bilello},
  \bibinfo{author}{M.~Rozycki}, \bibinfo{author}{J.~Kirby},
  \bibinfo{author}{J.~Freymann}, \bibinfo{author}{K.~Farahani},
  \bibinfo{author}{C.~Davatzikos},
\newblock \bibinfo{title}{Segmentation labels and radiomic features for the
  pre-operative scans of the {TCGA-LGG} collection},
\newblock \bibinfo{journal}{The Cancer Imaging Archive} \bibinfo{volume}{286}
  (\bibinfo{year}{2017}).
\bibitem[{Taghanaki et~al.(2019)Taghanaki, Zheng, Zhou, Georgescu, Sharma, Xu,
  Comaniciu, and Hamarneh}]{taghanaki2019combo}
\bibinfo{author}{S.~A. Taghanaki}, \bibinfo{author}{Y.~Zheng},
  \bibinfo{author}{S.~K. Zhou}, \bibinfo{author}{B.~Georgescu},
  \bibinfo{author}{P.~Sharma}, \bibinfo{author}{D.~Xu},
  \bibinfo{author}{D.~Comaniciu}, \bibinfo{author}{G.~Hamarneh},
\newblock \bibinfo{title}{Combo loss: Handling input and output imbalance in
  multi-organ segmentation},
\newblock \bibinfo{journal}{Computerized Medical Imaging and Graphics}
  \bibinfo{volume}{75} (\bibinfo{year}{2019}) \bibinfo{pages}{24--33}.
\bibitem[{Lin et~al.(2018)Lin, Goyal, Girshick, He, and Doll{\'a}r}]{lin99p}
\bibinfo{author}{T.-Y. Lin}, \bibinfo{author}{P.~Goyal},
  \bibinfo{author}{R.~Girshick}, \bibinfo{author}{K.~He},
  \bibinfo{author}{P.~Doll{\'a}r},
\newblock \bibinfo{title}{Focal loss for dense object detection},
\newblock \bibinfo{journal}{IEEE Transactions on Pattern Analysis and Machine
  Intelligence} \bibinfo{volume}{42} (\bibinfo{year}{2018})
  \bibinfo{pages}{318--327}.
\bibitem[{Paszke et~al.(2016)Paszke, Chaurasia, Kim, and
  Culurciello}]{paszke2016enet}
\bibinfo{author}{A.~Paszke}, \bibinfo{author}{A.~Chaurasia},
  \bibinfo{author}{S.~Kim}, \bibinfo{author}{E.~Culurciello},
\newblock \bibinfo{title}{Enet: A deep neural network architecture for
  real-time semantic segmentation},
\newblock \bibinfo{journal}{arXiv:1606.02147}  (\bibinfo{year}{2016}).
\bibitem[{Vo et~al.(2022)Vo, Dave, Bajpai, Kashef, and Khan}]{9861045}
\bibinfo{author}{T.~Vo}, \bibinfo{author}{P.~Dave},
  \bibinfo{author}{G.~Bajpai}, \bibinfo{author}{R.~Kashef},
  \bibinfo{author}{N.~Khan},
\newblock \bibinfo{title}{Brain tumor segmentation in {MRI} images using a
  modified {U-Net} model},
\newblock in: \bibinfo{booktitle}{2022 IEEE International Conference on Digital
  Health (ICDH)}, \bibinfo{year}{2022}, pp. \bibinfo{pages}{29--33}.
\bibitem[{Authors(2024)}]{tensorflowprofiler}
\bibinfo{author}{T.~Authors}, \bibinfo{title}{Profiler guide},
  \bibinfo{howpublished}{\url{https://www.tensorflow.org/guide/profiler}},
  \bibinfo{year}{2024}. \bibinfo{note}{Accessed: 2024-10-15}.
\bibitem[{Raza et~al.(2023)Raza, Bajwa, Mehmood, Anwar, and
  Jamal}]{raza2023dresu}
\bibinfo{author}{R.~Raza}, \bibinfo{author}{U.~I. Bajwa},
  \bibinfo{author}{Y.~Mehmood}, \bibinfo{author}{M.~W. Anwar},
  \bibinfo{author}{M.~H. Jamal},
\newblock \bibinfo{title}{{dResU-Net}: 3d deep residual {U-Net} based brain
  tumor segmentation from multimodal {MRI}},
\newblock \bibinfo{journal}{Biomedical Signal Processing and Control}
  \bibinfo{volume}{79} (\bibinfo{year}{2023}) \bibinfo{pages}{103861}.
\bibitem[{Ballestar and Vilaplana(2020)}]{ballestar2020brain}
\bibinfo{author}{L.~M. Ballestar}, \bibinfo{author}{V.~Vilaplana},
\newblock \bibinfo{title}{Brain tumor segmentation using {3D-CNNs} with
  uncertainty estimation},
\newblock \bibinfo{journal}{arXiv:2009.12188}  (\bibinfo{year}{2020}).
\bibitem[{Messaoudi et~al.(2021)Messaoudi, Belaid, Allaoui, Zetout, Allili,
  Tliba, Ben~Salem, and Conze}]{messaoudi2021efficient}
\bibinfo{author}{H.~Messaoudi}, \bibinfo{author}{A.~Belaid},
  \bibinfo{author}{M.~L. Allaoui}, \bibinfo{author}{A.~Zetout},
  \bibinfo{author}{M.~S. Allili}, \bibinfo{author}{S.~Tliba},
  \bibinfo{author}{D.~Ben~Salem}, \bibinfo{author}{P.-H. Conze},
\newblock \bibinfo{title}{Efficient embedding network for {3D} brain tumor
  segmentation},
\newblock in: \bibinfo{booktitle}{Brainlesion: Glioma, Multiple Sclerosis,
  Stroke and Traumatic Brain Injuries: 6th International Workshop, BrainLes
  2020, Part I 6}, \bibinfo{organization}{Springer}, \bibinfo{year}{2021}, pp.
  \bibinfo{pages}{252--262}.
\bibitem[{Tang et~al.(2021)Tang, Li, Shu, and Zhu}]{tang2021variational}
\bibinfo{author}{J.~Tang}, \bibinfo{author}{T.~Li}, \bibinfo{author}{H.~Shu},
  \bibinfo{author}{H.~Zhu},
\newblock \bibinfo{title}{Variational-autoencoder regularized {3D MultiResUNet}
  for the {BraTS} 2020 brain tumor segmentation},
\newblock in: \bibinfo{booktitle}{Brainlesion: Glioma, Multiple Sclerosis,
  Stroke and Traumatic Brain Injuries: 6th International Workshop, BrainLes
  2020, Part II 6}, \bibinfo{organization}{Springer}, \bibinfo{year}{2021}, pp.
  \bibinfo{pages}{431--440}.
\bibitem[{Peiris et~al.(2021)Peiris, Chen, Egan, and
  Harandi}]{peiris2021reciprocal}
\bibinfo{author}{H.~Peiris}, \bibinfo{author}{Z.~Chen},
  \bibinfo{author}{G.~Egan}, \bibinfo{author}{M.~Harandi},
\newblock \bibinfo{title}{Reciprocal adversarial learning for brain tumor
  segmentation: a solution to brats challenge 2021 segmentation task},
\newblock in: \bibinfo{booktitle}{International MICCAI Brainlesion Workshop},
  \bibinfo{organization}{Springer}, \bibinfo{year}{2021}, pp.
  \bibinfo{pages}{171--181}.
\bibitem[{Akbar et~al.(2021)Akbar, Fatichah, and Suciati}]{akbar2021unet3d}
\bibinfo{author}{A.~S. Akbar}, \bibinfo{author}{C.~Fatichah},
  \bibinfo{author}{N.~Suciati},
\newblock \bibinfo{title}{{Unet3D} with multiple atrous convolutions attention
  block for brain tumor segmentation},
\newblock in: \bibinfo{booktitle}{International MICCAI Brainlesion Workshop},
  \bibinfo{organization}{Springer}, \bibinfo{year}{2021}, pp.
  \bibinfo{pages}{182--193}.
\bibitem[{Jia et~al.(2021)Jia, Bai, Cai, Huang, and Xia}]{jia2021hnf}
\bibinfo{author}{H.~Jia}, \bibinfo{author}{C.~Bai}, \bibinfo{author}{W.~Cai},
  \bibinfo{author}{H.~Huang}, \bibinfo{author}{Y.~Xia},
\newblock \bibinfo{title}{{HNF-NetV2} for brain tumor segmentation using
  multi-modal {MR} imaging},
\newblock in: \bibinfo{booktitle}{International MICCAI Brainlesion Workshop},
  \bibinfo{organization}{Springer}, \bibinfo{year}{2021}, pp.
  \bibinfo{pages}{106--115}.
\bibitem[{Li et~al.(2021)Li, Shen, Wen, He, and Pan}]{li2021automatic}
\bibinfo{author}{Z.~Li}, \bibinfo{author}{Z.~Shen}, \bibinfo{author}{J.~Wen},
  \bibinfo{author}{T.~He}, \bibinfo{author}{L.~Pan},
\newblock \bibinfo{title}{Automatic brain tumor segmentation using multi-scale
  features and attention mechanism},
\newblock in: \bibinfo{booktitle}{International MICCAI Brainlesion Workshop},
  \bibinfo{organization}{Springer}, \bibinfo{year}{2021}, pp.
  \bibinfo{pages}{216--226}.
\bibitem[{Ma and Chen(2021)}]{ma2021nnunet}
\bibinfo{author}{J.~Ma}, \bibinfo{author}{J.~Chen},
\newblock \bibinfo{title}{{NnUNet} with region-based training and loss
  ensembles for brain tumor segmentation},
\newblock in: \bibinfo{booktitle}{International MICCAI Brainlesion Workshop},
  \bibinfo{organization}{Springer}, \bibinfo{year}{2021}, pp.
  \bibinfo{pages}{421--430}.
\bibitem[{Hatamizadeh et~al.(2021)Hatamizadeh, Nath, Tang, Yang, Roth, and
  Xu}]{hatamizadeh2021swin}
\bibinfo{author}{A.~Hatamizadeh}, \bibinfo{author}{V.~Nath},
  \bibinfo{author}{Y.~Tang}, \bibinfo{author}{D.~Yang}, \bibinfo{author}{H.~R.
  Roth}, \bibinfo{author}{D.~Xu},
\newblock \bibinfo{title}{{Swin UNTER:} swin transformers for semantic
  segmentation of brain tumors in {MRI} images},
\newblock in: \bibinfo{booktitle}{International MICCAI Brainlesion Workshop},
  \bibinfo{organization}{Springer}, \bibinfo{year}{2021}, pp.
  \bibinfo{pages}{272--284}.
\bibitem[{Roth et~al.(2021)Roth, Keller, Franke, Neumuth, and
  Schneider}]{roth2021multi}
\bibinfo{author}{J.~Roth}, \bibinfo{author}{J.~Keller},
  \bibinfo{author}{S.~Franke}, \bibinfo{author}{T.~Neumuth},
  \bibinfo{author}{D.~Schneider},
\newblock \bibinfo{title}{Multi-plane {UNet++} ensemble for glioblastoma
  segmentation},
\newblock in: \bibinfo{booktitle}{International MICCAI Brainlesion Workshop},
  \bibinfo{organization}{Springer}, \bibinfo{year}{2021}, pp.
  \bibinfo{pages}{285--294}.
\bibitem[{Alwadee et~al.(2024)Alwadee, Sun, Qin, and Langbein}]{alwadee2024}
\bibinfo{author}{E.~Alwadee}, \bibinfo{author}{X.~Sun},
  \bibinfo{author}{Y.~Qin}, \bibinfo{author}{F.~Langbein},
\newblock \bibinfo{title}{Assessing and enhancing the robustness of brain tumor
  segmentation using a probabilistic deep learning architecture},
\newblock in: \bibinfo{booktitle}{ISMRM 2024 Conference Proceedings},
  \bibinfo{organization}{International Society for Magnetic Resonance in
  Medicine (ISMRM)}, \bibinfo{year}{2024}. \URLprefix
  \url{https://submissions.mirasmart.com/ISMRM2024/Itinerary/ConferenceMatrixEventDetail.aspx?ses=D-173},
  \bibinfo{note}{abstract 4526, Session 47}.

\end{thebibliography}

\vbox{\bio{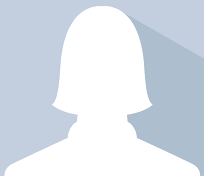}
\noindent {\bf Ebtihal J. Alwadee} received her BSc (Hons) in Information Systems from King Khalid University, Saudi Arabia, in 2013 and her MSc in Computer Science from California State University, Fullerton, USA, in 2020. She is currently a PhD candidate at Cardiff University, with research interests in visual computing, healthcare, deep learning, medical image segmentation, and explainable AI.
\endbio}

\vbox{\bio{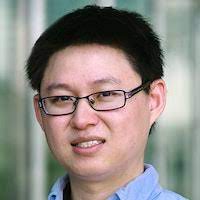}
\noindent {\bf Yipeng Qin} received a BSc degree in electrical engineering from Shanghai Jiao Tong University, China, and a PhD degree from the National Centre for Computer Animation (NCCA), Bournemouth University, UK. He was a Postdoctoral Research Fellow with the Visual Computing Center (VCC), King Abdullah University of Science and Technology (KAUST), Saudi Arabia. He is currently a Lecturer at the School of Computer Science and Informatics, Cardiff University, UK. His current research interests include deep learning, computer vision, computer graphics, and human–computer interaction (HCI), with a focus on generative modeling and visual content creation.
\endbio}

\vbox{\bio{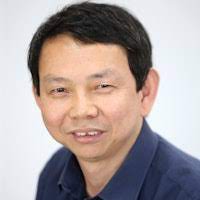}
\noindent {\bf Xianfang Sun} received a PhD degree from the Institute of Automation, Chinese Academy of Sciences, in 1994. He is currently a Senior Lecturer at the School of Computer Science, Cardiff University, UK. His main research interests include computer vision, computer graphics, pattern recognition, and artificial intelligence.
\endbio}

\vbox{\bio{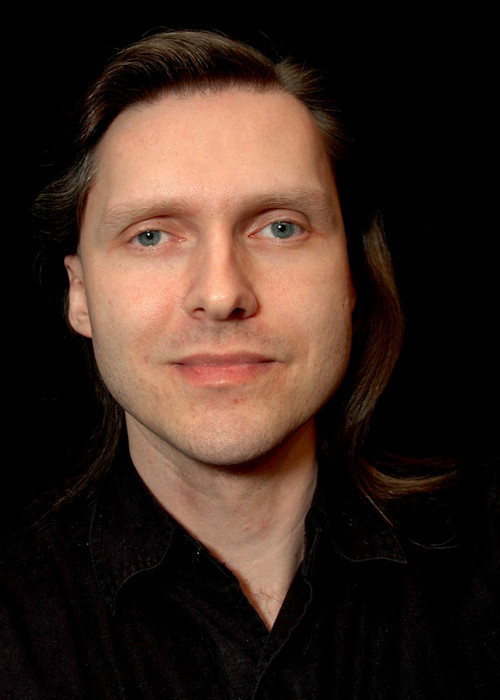}
\noindent {\bf Frank C.\ Langbein} received his Mathematics degree from Stuttgart University, Germany, in 1998 and a PhD from Cardiff University, UK, in 2003. He is currently a senior lecturer at the School of Computer Science and Informatics, Cardiff University, where he is a member of the Visual Computing Research Section and leads the Quantum Control Research Group. He co-leads Qyber, a research network in quantum control, geometry, medical diagnosis, and machine learning. His research interests include control, machine learning, and geometry applied in quantum technologies, visual computing, geometric modeling, and healthcare. He is a member of the Institute of Electrical and Electronics Engineers (IEEE) and the American Mathematical Society (AMS).
\endbio}

\newpage

\centerline{\LARGE Supplementary Material}

\section{Investigation of the Effect of Different Upsampling Approaches}

In order to find the optimal upsampling method for the LATUP-Net architecture, we investigate two distinct techniques to identify the one that yields the best performance. Common upsampling approaches in lightweight networks are linear interpolation and transposed convolution, which are both tested here. Since our dataset consists of 3D MRI scans, trilinear interpolation is essential as it enables estimation across three dimensions by interpolating the values between neighboring voxels in a continuous manner. Trilinear interpolation preserves spatial relationships while performing upsampling and ensures that interpolated values smoothly transition between the feature mappings in the previous layer. Transposed convolution, instead, works by reversing the process of convolution to upsample the input, with the output size controlled by the kernel and stride parameters. Although transposed convolution can recover spatial resolution, it tends to introduce checkerboard artifacts, which can negatively impact segmentation performance. The experimental outcomes presented in Table~S1 demonstrate that trilinear interpolation outperforms transposed convolution across all Dice scores, particularly in the tumor core (TC) and enhancing tumor (ET) regions. This demonstrates that, while both methods maintain same parameter efficiency, trilinear interpolation offers better precision in capturing finer details of tumor structures, making it the preferred upsampling strategy for LATUP-Net.

\section{Comparison of Different Attention Mechanisms}

We analyze the efficiency of various attention mechanisms in brain tumor segmentation on the BraTS~2020 test set. The evaluation is based on the Dice similarity coefficient (DSC) for whole tumor (WT), tumor core (TC), and enhancing tumor (ET) segmentation. Additionally, we provide the total training time for each model for reference, although note this information dependends strongly on the GPU hardware used (an RTX3060 in this case). Thus, model complexity comparisons based on GFLOPs, the number of parameters, and inference time are provided in Section~5.3, Table~5 of the main paper.

\section{Investigation of the Effect of Different Loss Functions}

In the LATUP-Net architecture, we evaluated multiple loss functions to determine the most effective one for optimizing the segmentation of 3D MRI data. The choice of loss function is critical as it directly influences the network’s ability to handle class imbalance and predict accurate segmentation. We experimented with several loss functions: Dice loss, Focal loss, Wasserstein Dice Loss, a combined Dice with Focal loss, weighted Dice loss with different weights for each class (WT, TC, ET), and weighted Dice loss with equal weights across all classes (WT, TC, ET).

The results summarized in Table~S3 demonstrate that while each loss function has its strengths, the proposed weighted Dice loss (with class-specific weights) yields the best segmentation results.

\section{Investigation of the Effect of Different Dropout Rates}

Dropout is a widely used regularization technique that prevents overfitting by randomly dropping neurons during training, reducing reliance on any specific features and forcing the network to generalize better. We tested three dropout rates: $0.2$, $0.3$, and $0.4$.

As shown in Table~S4, a dropout rate of $0.2$ yielded the best performance, providing a balance between reducing overfitting and maintaining model capacity. The dropout rates of $0.3$ and $0.4$ performed slightly worse, likely due to too many neurons being dropped, leading to underfitting.

\begin{table*}[t]
\RaggedRight
\textbf{Table~S1}\\
Segmentation performance results for employing different upsampling strategies in LATUP-Net on the BraTS~2020 test set. WT -- whole tumor, TC -- tumor core, ET -- enhancing tumor. Experiments were conducted using an RTX3060 GPU, after adding the proposed PC block, and the model was trained using the Dice loss function.\\[1ex]
\centering
\begin{tblr}{
  cell{1}{1} = {r=2}{},
  cell{1}{2} = {r=2}{},
  cell{1}{3} = {c=3}{halign=c},
  hline{1,3-5} = {-}{},
  hline{2} = {3-5}{},
}
Upsampling Strategy     & Number of parameters & Dice Score  &             &             \\
                        &                      & \textbf{WT} & \textbf{TC} & \textbf{ET} \\
Convolution Transpose   & $3,069,060$          & $0.8784$    & $0.8338$    & $0.7085$    \\
Trilinear interpolation & $3,069,060$          & $0.8823$    & $0.8570$    & $0.7150$    \\
\end{tblr}
\end{table*}

\begin{table*}[t]
\RaggedRight
\textbf{Table~S2}\\
Efficiency analysis of various attention mechanisms for brain tumor segmentation on the BraTS~2020 test set. WT -- whole tumor, TC -- tumor core, ET -- enhancing tumor. The table includes the Dice score and total training time for each model. Experiments were conducted using an RTX3060 GPU, after adding the proposed PC block, and the model was trained using the Dice loss function.\\[1ex]
\centering
\begin{tblr}{
  cell{1}{1} = {r=2}{},
  cell{1}{2} = {c=3}{halign=c},
  hline{1,3} = {-}{},
  hline{2} = {2-4}{},
}
Attention Mechanism  & Dice Score  &             &             \\
                     & \textbf{WT} & \textbf{TC} & \textbf{ET} & Total Training Time (hours) \\
PC+SE                & $88.52$     & $83.26$     & $\mathbf{71.86}$ & $13.20$  \\
PC+CBAM              & $89.38$     & $\mathbf{84.36}$ & $70.01$     & $18.22$  \\
PC+CBAM+SE           & $88.91$     & $83.87$     & $70.25$     & $16.80$  \\
PC+(SE-3D)           & $89.05$     & $83.91$     & $69.83$     & $13.89$  \\
PC+Residual SE       & $\mathbf{89.60}$ & $84.06$     & $70.79$     & $15.89$  \\
PC+ECA               & $84.47$     & $80.12$     & $63.19$     & $15.94$  \\\hline
\end{tblr}
\end{table*}

\begin{table*}[t]
\RaggedRight
\textbf{Table~S3}\\
Segmentation performance results for employing different loss functions in LATUP-Net on the BraTS 2020 test set. WT -- whole tumor, TC -- tumor core, ET -- enhancing tumor.  Experiments were conducted using an RTX3060 GPU, after adding the proposed PC block.\\[1ex]
\centering
\begin{tblr}{
  cell{1}{1} = {r=2}{},
  cell{1}{2} = {c=3}{halign=c},
  hline{1,3} = {-}{},
  hline{2} = {2-4}{},
}
Loss Function                        & Dice Score  &             &             \\
                                     & \textbf{WT} & \textbf{TC} & \textbf{ET} \\
Dice Loss                            & $88.52$     & $83.26$     & $71.86$     \\
Focal Loss                           & $87.15$     & $81.34$     & $70.21$     \\
Wasserstein Dice Loss                     & $87.42$     & $81.76$     & $70.94$     \\
Dice + Focal Loss                    & $88.63$     & $83.84$     & $72.19$     \\
Weighted Dice Loss (Different Weights)& $88.68$     & $84.12$     & $73.56$     \\
Weighted Dice Loss (Equal Weights)    & $88.61$     & $83.72$     & $72.98$     \\
The proposed WDL                     & $88.72$     & $84.71$     & $74.49$     \\\hline
\end{tblr}
\end{table*}

\section{Investigation of the Effect of Different Regularizers}

Regularization is another important technique to prevent overfitting. We experimented with L1, L2, and a combination of both L1 and L2 (Elastic Net) regularization. L1 regularization encourages sparsity, while L2 regularization penalizes large weights. We tested different values for each type of regularization: L1 with $0.1$, $0.2$ weights, and L2 with $0.1$, $0.2$ weights, and L1 + L2 with weights $(0.1, 0.2)$.

As summarized in Table~S5, L2 regularization with a weight of $0.2$ resulted in the best segmentation performance, providing significant benefits in terms of preventing overfitting, as observed in the loss curve. The combination of L1 and L2 regularization with values $(0.1, 0.2)$ also performed well, while stronger L1 regularization with $0.2$ weight reduced performance slightly.

\begin{table*}[t]
\RaggedRight
\textbf{Table~S4}\\
Segmentation performance results for employing different dropout rates in LATUP-Net on the BraTS~2020 test set. WT -- whole tumor, TC -- tumor core, ET -- enhancing tumor. Experiments were conducted using an RTX3060 GPU, after adding the proposed PC block, and the model was trained using the Dice loss function.\\[1ex]
\centering
\begin{tblr}{
  cell{1}{1} = {r=2}{},
  cell{1}{2} = {c=3}{halign=c},
  hline{1,3} = {-}{},
  hline{2} = {2-4}{},
}
Dropout Rate   & Dice Score  &             &             \\
               & \textbf{WT} & \textbf{TC} & \textbf{ET} \\
$0.2$          & $89.25$     & $85.19$     & $71.23$     \\
$0.3$          & $86.23$     & $83.70$     & $67.05$     \\
$0.4$          & $85.55$     & $83.05$     & $66.21$     \\\hline
\end{tblr}
\end{table*}

\begin{table*}[t]
\RaggedRight
\textbf{Table~S5}\\
Segmentation performance results for employing different regularizers in LATUP-Net on the BraTS~2020 test set. WT -- whole tumor, TC -- tumor core, ET -- enhancing tumor. Experiments were conducted using an RTX3060 GPU, after adding the proposed PC block, and the model was trained using the Dice loss function.\\[1ex]
\centering
\begin{tblr}{
  cell{1}{1} = {r=2}{},
  cell{1}{2} = {c=3}{halign=c},
  hline{1,3} = {-}{},
  hline{2} = {2-4}{},
}
Regularizer, weight   & Dice Score  &             &             \\
                      & \textbf{WT} & \textbf{TC} & \textbf{ET} \\
L1, $0.1$             & $85.72$     & $83.29$     & $66.62$     \\
L1, $0.2$             & $85.35$     & $82.85$     & $66.13$     \\
L2, $0.1$             & $86.14$     & $83.65$     & $66.91$     \\
L2, $0.2$             & $88.01$     & $83.14$     & $70.23$     \\
(L1,L2), $(0.1, 0.2)$ & $86.23$     & $83.70$     & $67.05$     \\\hline
\end{tblr}
\vspace{10in}
\end{table*}

\end{document}